%% file: paperI.tex
\documentclass[preprint2]{aastex}
\usepackage{indentfirst} 
\usepackage{amsmath}
\usepackage{natbib}

\shorttitle{Dual-Frequency Observations of ISS}
\shortauthors{Koay et al.}

\title{Dual-Frequency Observations of 140 Compact, Flat-Spectrum Active Galactic Nuclei for Scintillation-Induced Variability\\}
\author{J. Y. Koay\altaffilmark{1*}, J.-P. Macquart\altaffilmark{1}, B. J. Rickett\altaffilmark{2}, H. E. Bignall\altaffilmark{1}, J. E. J. Lovell\altaffilmark{3}, C. Reynolds\altaffilmark{1}, D. L. Jauncey\altaffilmark{4,5}, T. Pursimo\altaffilmark{6}, L. Kedziora-Chudczer\altaffilmark{7}, and R. Ojha\altaffilmark{8,9}}

\altaffiltext{*}{e-mail: kevin.koay@icrar.org}

\altaffiltext{1}{International Centre for Radio Astronomy Research, Curtin University, Bentley, WA 6102, Australia}

\altaffiltext{2}{Department of Electrical and Computer Engineering, University of California, San Diego, La Jolla, CA 92093, USA}

\altaffiltext{3}{School of Mathematics and Physics, University of Tasmania, TAS 7001, Australia}

\altaffiltext{4}{CSIRO Astronomy and Space Science, Australia Telescope National Facility, Epping, NSW 1710, Australia}

\altaffiltext{5}{Mount Stromlo Observatory, Weston, ACT 2611, Australia}

\altaffiltext{6}{Nordic Optical Telescope, Apartado 474, 38700 Santa Cruz de La Palma, Spain}

\altaffiltext{7}{School of Physics and Astrophysics, University of New South Wales, Sydney, NSW 2052, Australia}

\altaffiltext{8}{NASA, Goddard Space Flight Center, Greenbelt, MD 20771, USA}

\altaffiltext{9}{Institute for Astrophysics \& Computational Sciences, The Catholic University of America, 620
Michigan Ave., N.E., Washington, DC 20064, USA}

\clearpage

   \begin{abstract}
The 4.9 GHz Micro-Arcsecond Scintillation-Induced Variability (MASIV) Survey detected a drop in Interstellar Scintillation (ISS) for sources at redshifts $z \gtrsim 2$, indicating an apparent increase in angular diameter or a decrease in flux density of the most compact components of these sources, relative to their extended emission. This can result from intrinsic source size effects or scatter broadening in the Intergalactic Medium (IGM), in excess of the expected $(1+z)^{1/2}$ angular diameter scaling of brightness temperature limited sources resulting from cosmological expansion. We report here 4.9 GHz and 8.4 GHz observations and data analysis for a sample of 140 compact, flat-spectrum sources which may allow us to determine the origin of this angular diameter-redshift relation by exploiting their different wavelength dependences. In addition to using ISS as a cosmological probe, the observations provide additional insight into source morphologies and the characteristics of ISS. As in the MASIV Survey, the variability of the sources is found to be significantly correlated with line-of-sight H$\alpha$ intensities, confirming its link with ISS. For 25 sources, time delays of about 0.15 to 3 days are observed between the scintillation patterns at both frequencies, interpreted as being caused by a shift in core positions when probed at different optical depths. Significant correlation is found between ISS amplitudes and source spectral index; in particular, a large drop in ISS amplitudes is observed at $\alpha < -0.4$ confirming that steep spectrum sources scintillate less. We detect a weakened redshift dependence of ISS at 8.4 GHz over that at 4.9 GHz, with the mean variance at 4-day timescales reduced by a factor of 1.8 in the $z > 2$ sources relative to the $z < 2$ sources, as opposed to the factor of 3 decrease observed at 4.9 GHz. This suggests scatter broadening in the IGM, but the interpretation is complicated by subtle selection effects that will be explored further in a follow-up paper.  
   \end{abstract}
   

\begin{document}

   \maketitle
   
   \section{Introduction}\label{introduction}
   
It is well established that the intraday variability (IDV) observed in many compact, flat-spectrum active galactic nuclei (AGN) at centimetre wavelengths is predominantly caused by scintillation in the turbulent and ionized interstellar medium (ISM) of our own Galaxy. The idea was first proposed by \citet{heeschenrickett87} in order to resolve the brightness temperature problem in AGN, where intrinsic variability on the time-scales observed implied brightness temperatures well over the $10^{12}$ K inverse-Compton limit for incoherent synchrotron emission \citep{hunstead72,heeschen84}. Substantial observational evidence has accumulated in the last decade to support this theory. Time delays of up to 8 minutes have been observed in the scintillation patterns of the most rapid scintillators at widely spaced telescopes \citep{jaunceyetal00, dennett-thorpedebruyn02, bignalletal06}, as would be expected of interference patterns drifting across the surface of the Earth as a result of relative motion between the ISM and the Earth. Annual cycles have also been detected in AGN variability time-scales \citep{rickettetal01, jaunceymacquart01, bignalletal03, dennett-thorpedebruyn03, jaunceyetal03}, interpreted as being modulated by the orbital motion of the Earth around the Sun. When the Earth's motion is parallel to the motion of the scattering medium, the variability time-scales are longer, while shorter timescale variability occurs when the Earth's motion is anti-parallel to that of the scattering medium.

More recently, the Microarcsecond Scintillation-Induced Variability (MASIV) Survey \citep{lovelletal03, lovelletal08} provided further confirmation of interstellar scintillation (ISS) as the principal mechanism behind IDV. In a 4.9 GHz survey of more than 500 compact, flat-spectrum sources at 4 separate epochs spaced throughout a year, it was found that more than half of the sources were variable in at least one epoch. The survey showed a strong correlation between AGN variability and Galactic latitudes, as well as line of sight H$\alpha$ intensity (as an estimate of the emission measure of the ISM), thus strengthening the link between IDV phenomena and the ionized ISM.

The effects of scintillation are highly dependent on the angular size of the source. Compact sources tend to scintillate more than extended sources, analogous to the fact that `stars twinkle, but planets do not' when observed through the Earth's atmosphere. For weak interstellar scattering (WISS), which is usually the case for scintillating sources at frequencies $\geq$ 5 GHz at mid-Galactic latitudes, the angular size of the source must be comparable to or smaller than the size of the first Fresnel zone \citep{narayan92} at the scattering screen. The variations in the flux density of the MASIV sources indicate that a significant portion of the emission comes from compact components with angular diameters on a scale of 10 to 50 microarseconds. Follow-up observations of the morphologies of scintillating and non-scintillating sources using Very Long Baseline Interferometry (VLBI) confirm that, at milliarcsecond scales, the scintillating sources are more core-dominated than non-scintillating sources \citep{ojhaetal04a, ojhaetal04b}, where a large proportion of their flux densities are confined in an ultra-compact core region, best interpreted as highly Doppler-boosted jet emission with intrinsic brightness temperatures close to the inverse-Compton limit of $10^{12}$ K. In the presence of stronger, milliarcsecond-scale jets, scintillation effects are diminished.

A most tantalizing result of the MASIV survey was the discovery of a significant drop in the fraction of scintillating sources, as well as their variability amplitudes, at redshifts above 2 \citep{lovelletal08}. While the angular diameters of a population of sources limited by a maximum brightness temperature are expected to scale with $(1+z)^{1/2}$ in an expanding $\Lambda$CDM Universe \citep{rickettetal07}, due to the source brightness temperature in the observer's frame being a factor of $(1+z)$ lower relative to the source brightness temperature in the comoving frame, the redshift dependence observed in the survey was found to be in excess of this effect. Therefore, it can be attributed either to an additional increase in typical source angular diameters, or a decrease in flux densities of the aforementioned ultra-compact components of the AGN relative to their more extended components. An increase in the angular diameter can be a result of angular broadening as the radio waves propagate through the turbulent, ionized intergalactic medium (IGM), or a decrease in the Doppler boosting factor in AGN jets at earlier epochs. A reduction in flux density of the ultra-compact component can also be due to a decrease in the Doppler boosting factor of the jets or even a decrease in the prevalence of such very compact objects at higher redshifts.

Identifying with certainty the cause of this redshift dependence in AGN ISS has profound cosmological implications. If the cause is intrinsic to the sources, it provides new results on the evolution of AGN morphologies at scales two orders of magnitude finer than that available to VLBI. On the other hand, detection of angular broadening in the IGM would provide a new observational tool for the study of the majority of baryons (90\% are thought to reside in the IGM, see e.g. \citet{fukugitapeebles04}) and their evolution after the epoch of reionization. Being sensitive to turbulent, ionized components of the IGM, it will thus complement Ly$\alpha$ studies of the IGM which are sensitive only to the neutral component, as well as UV and X-ray observations of the hot intra-cluster medium (ICM). Even more exciting is the prospect of detecting the elusive warm-hot intergalactic medium (WHIM), predicted by cosmological hydrodynamical simulations \citep{cenostriker99, daveetal01, cenostriker06} to exist in an almost fully ionized state at temperatures of $10^{5}$ to $10^{7}$ K due to gravitational shock heating. The WHIM  has so far been very difficult to detect conclusively \citep{bregman07}, and forms one of the key science drivers of the next generation of X-ray instruments.

To determine the origin of this redshift dependence in ISS, we have conducted multi-frequency observations of a sub-sample of the MASIV sources using the Very Large Array (VLA). Observing at multiple frequencies can potentially provide an effective technique for discriminating the cause of the redshift dependence in AGN scintillation from among the three possible explanations - cosmological expansion, angular broadening in the IGM and evolution of AGN jets. Scatter broadening in the ionized IGM should have a stronger wavelength dependence compared to the angular size-wavelength dependence of the source core, leading to a decrease in the redshift scaling at higher frequencies. These observations, therefore, have the ultimate goal of using ISS as a cosmological probe - potentially of AGN jet evolution, turbulence in the IGM, or the curvature of the Universe.

While determining the cause of the redshift dependence of ISS remains the main objective of the study, the opportunity to gain additional insight into ISS phenomena and AGN morphology provides additional motivations for conducting the observations. Firstly, the 11-day duration of these observations (as opposed to the 3 or 4 day epochs in the MASIV survey) gives improved constraints on source scintillation timescales. Secondly, multi-frequency observations of ISS provide a means of detecting any angular offset in the positions of the AGN cores at different frequencies, observed as a delay in the scintillation patterns between frequencies as the scattering screen drifts across the source, as seen in PKS 1257--326 \citep{bignalletal03}. Such frequency core-shifts have also been observed in VLBI studies \citep{kovalevetal08}, and are interpreted in terms of opacity effects in the source jet. Thirdly, multi-frequency observations also enable us to estimate the instantaneous spectral indices of the sources based on concurrent mean flux densities to study its relationship to ISS (further details in Section~\ref{spectralindexsec}). Finally, together with the MASIV Survey, the experiment enables us to place a lower limit on the detectability of ISS amongst the presence of noise and other systematic errors using the VLA, in addition to providing a platform for exploring various methods of estimating and accounting for these errors. These observations thus act as a demonstrator for future large scale surveys such as the planned Australian Square Kilometre Array Pathfinder (ASKAP) Variables and Slow Transients (VAST) Survey \citep{murphyandchatterjee09}.

Section~\ref{observations} of this paper describes the observations and data reduction process for this follow-up to MASIV. This is followed by a detailed elucidation of the various methods used in the analysis of the data (including error estimation and correction), along with the results (Section~\ref{analysisresults}). Section~\ref{conclusion} presents our conclusions. Further interpretation of these results with regards to the redshift dependence of ISS in AGN and its cosmological implications is discussed in a separate paper, which will investigate the possible source selection effects that may lead to biases in the interpretation.
   
   \section{Observations, Data Reduction and Calibration}\label{observations} 

A sample of 140 sources were selected from the original MASIV set of sources. 70 of these sources have measured redshifts of $z > 2$, while the remaining 70 have a redshift of $z < 2$ as a control sample. Care was taken to ensure that both groups have similar distribution in terms of Galactic latitudes and that both have equal proportions of sources with weak ($<$ 0.3 Jy) and strong ($>$ 0.3 Jy) flux densities to avoid source selection biases related to these factors. While it is obvious that ISS has a dependence on Galactic latitudes, the MASIV survey found that low flux density sources tend to scintillate more than strong sources, consistent with a brightness temperature limited sample of sources. The sources that were selected were expected to have flux densities above 100 mJy at 8.4 GHz, and to be unresolved when observed with the VLA in its largest configuration (maximum baseline of 36.4 km and a FWHM synthesized beamwidth of 0.24 arcsecond at 8.4 GHz). A list of these sources can be found in Appendix A together with their corresponsing observed flux densities, spectral indices and variability characteristics.

Observations were carried out over 11 days from 2009 January 15 to 2009 January 25 using the VLA. The instrument was divided into 2 subarrays. One subarray comprised of 14 EVLA antennas observing with two 50 MHz IF channels in continuum mode, one centered at 4.9 GHz (6 cm) and another at 6.6 GHz (4.5 cm). The second subarray was a mix of 13 VLA and EVLA antennas observing at a centre frequency of 8.4 GHz (3.6 cm) with two continuum mode IF channels (contiguous 50 MHz bandwidths). During the observations, each source was observed for 1 minute at $\approx 2$ hour intervals simultaneously on both subarrays. The correlator integration time was set to 3.3 seconds. Observations of these target sources were interspersed with observations of the primary calibrator (3C286) and 23 secondary calibrators, selected from the list of sources in the VLA calibrator manual.

Unfortunately, 12 of the antennas from the 4.9 and 6.6 GHz subarray, as well as 8 antennas from the 8.4 GHz subarray, encountered data losses on the 7th and 8th day of the observations due to failure in the optical fiber links. This left only a single baseline on the 4.9 GHz subarray, which had to be flagged, and 10 baselines on the 8.4 GHz subarray. Thus no data was obtained at 4.9 GHz and 6.6 GHz on those days, while data at 8.4 GHz was retained, though with a reduced number of baselines. The observations were conducted during reconfiguration of the VLA between the BnA and B configurations, so recently moved antennas may have introduced pointing errors into the data. These were corrected for as much as possible via careful calibration. All data from 2 antennas in the 8.4 GHz subarray in which the pointing errors were the worst were removed entirely.

The data was loaded into the AIPS software package \citep{greisen03} using the task FILLM, which by default corrects for known antenna gain-elevation dependence. Upon inspection of the raw data, it was found that the 6.6 GHz data were subjected to extensive contamination by radio frequency interference (RFI). Hence, they are excluded from the present study and from the discussions that follow. There were also large increases in the amplitude variations in the uncalibrated data from day 7 of the observations onwards (typically increasing from 1\% to 4\% rms variations), after the technical problems were encountered on the VLA. These variations were attributed to system gain variations. Although our calibration successfully removed most of the effects, some residuals remain. These residuals are larger than the residuals in the first 6 days of continuous observations when the system gains were more stable. Therefore, the data between day 7 and 11 were treated with extra caution. Discarding all the data after 6 days may reduce the errors due to possible false variability, but results in a dataset with a reduced timespan with higher statistical uncertainties in the estimation of the variability characteristics. As a compromise, all subsequent data analyses were carried out using both sets of data --- one using data only from the first 6 days, and another using data from the entire duration of the observations from which comparisons could be made. This provided another means of cross-examining the results of our analysis.

Standard techniques were used to calibrate for atmospheric effects, as well as antenna gain and pointing errors, using the secondary calibrators. Phase self-calibration was then applied to all the target sources. Polarization calibration and parallactic angle corrections were also applied. After calibration, each of the target sources were examined for outlying points and spurious data, which were then flagged. The data was then converted into FITS format, so that they could be loaded into the Miriad software package \citep{saultetal95} which provides a convenient means of generating the desired output in plain text format. Using Miriad, the visibilities were coherently averaged over 1 minute and over all baselines (as well as across both channels for the 8.4 GHz data) to produce the lightcurves for each source.

It was essential to ensure that the secondary calibrators were not themselves variable down to the sub 0.5\% variability levels probed by the survey. We inspected the target source lightcurves by eye for possible contamination by spurious variability in the secondary calibrators by looking for similar variability patterns in sources that had been calibrated using the same calibrator. Such patterns would be particularly obvious for the stronger sources where calibration errors are expected to dominate over errors due to random noise. While no calibrators were found to be variable this way, we cannot rule out the presence of calibrator variability that is undetectable by eye, as they will probably be superposed on top of real scintillation or other sources of errors. This preliminary method of detecting calibrator variability was thus supplemented, and its effects corrected for, with further, more quantitative techniques discussed in Section~\ref{errors}. No recalibration of the target sources was necessary as these errors were accounted for via subtraction of the estimated error values from the calculated variability amplitudes for each source.

The lightcurves were then examined for repeating daily variations indicating confusion due to extended structure or contaminating sources close by. The observations were scheduled in sidereal time and each source was observed at the same time each sidereal day. Therefore, any confusion or resolution effects appear as repeating patterns with a 1 day period, with the amplitude of the variations being independent of the source flux density. Such repeating patterns can also result from residual gain and pointing errors from the calibration process, in which case the apparent variations will be a percentage of the source flux density. Clues to this false variability can also appear in the structure functions of the source (to be explained further in the next section) at integer multiples of a sidereal day. This process found that slightly more than a third of the sources displayed such daily repeating patterns on at least one frequency, some of them superposed on top of larger variations. For such sources, snapshot images and plots of the visibility amplitudes vs. uv-distances were produced to determine if the presence of structure or confusing sources could be confirmed. Only 3 of these sources were found to show resolution effects (and were subsequently remedied by the removal of the longer baselines), particularly at 8.4 GHz. There was also no clear evidence of contaminating sources nearby any of our sources. However, the daily repeating patterns for the vast majority of the remaining sources with such problems (about 95\% of them) turned out to be residual gain errors and pointing errors. This conclusion was arrived at after it was found that these patterns which repeat daily (typically $\lesssim 1\%$ rms variations) were almost always found on the higher flux density, low-variability sources. It is in such sources where calibration errors are expected to dominate. Further details on the estimation and correction of these errors is presented in Section~\ref{errors}.    
      
   \section{Data Analysis and Results}\label{analysisresults}  
      
      \subsection{Lightcurves and Structure Functions}\label{sf}
\begin{figure*}[!htp]
     \begin{center}
     \includegraphics[scale=0.6]{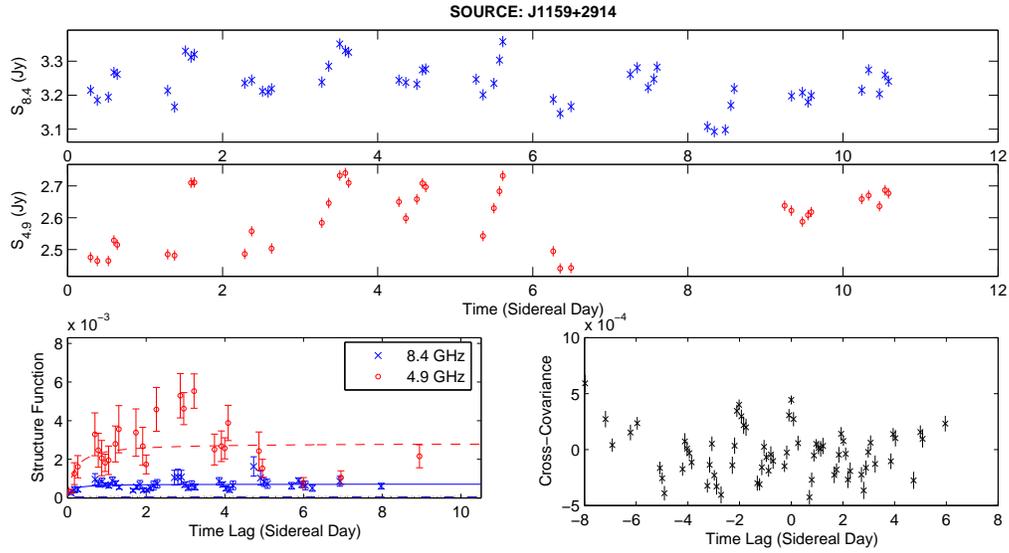}
     \end{center}
     \caption{{Lightcurves for the source J1159+2914 at 8.4 GHz (top) and 4.9 GHz (middle), with their corresponding structure functions (bottom left, where the solid curve and dashed curve represent the model fits at 8.4 GHz and 4.9 GHz respectively, the dash-dot line represents $D_{noise}$ at 4.9 GHz and the dotted line represents $D_{noise}$ at 8.4 GHz) and cross-covariance function (bottom right).} \label{11982923}}
     \end{figure*}

\begin{figure*}[!htp]
     \begin{center}
     \includegraphics[scale=0.6]{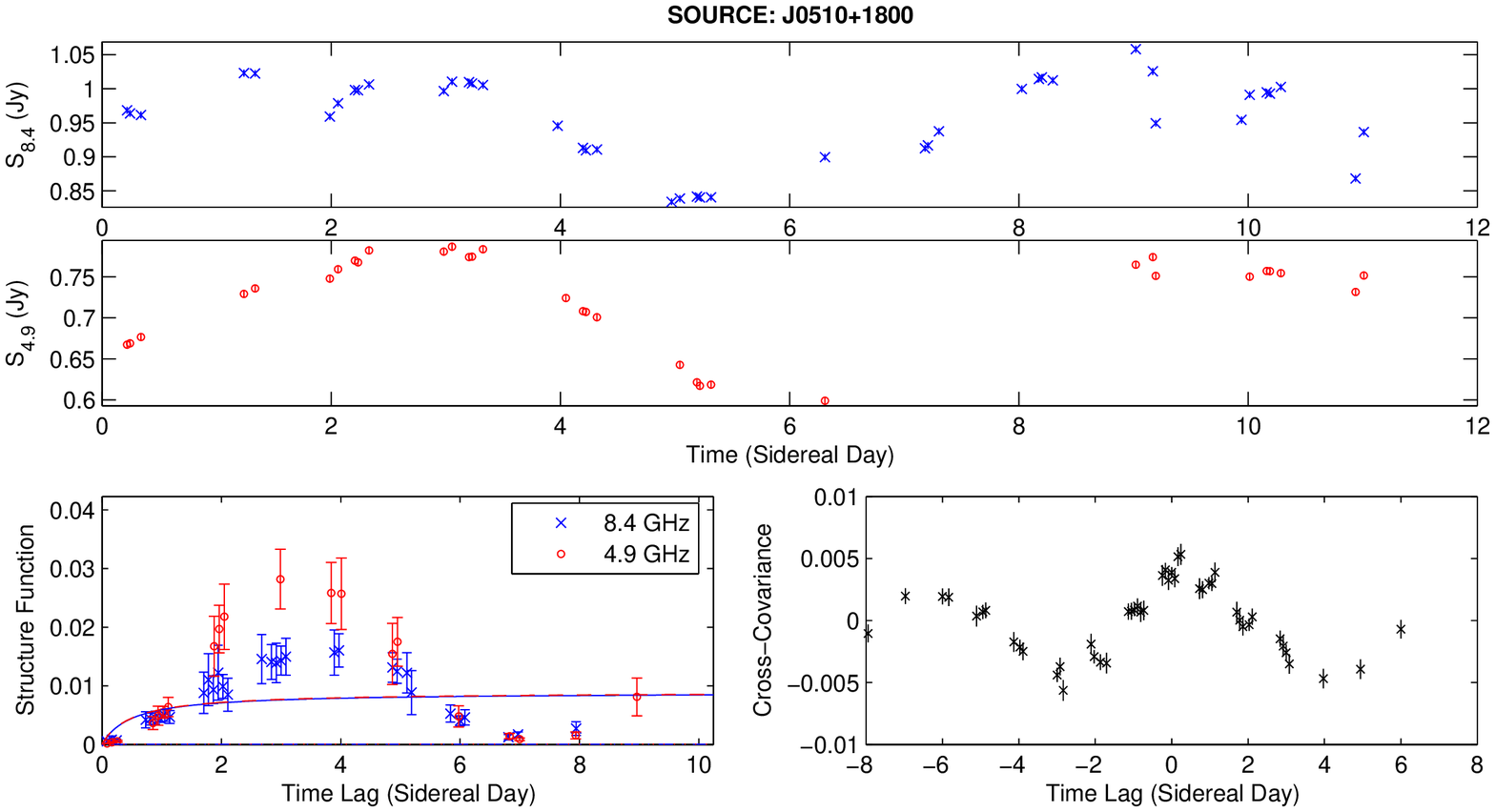}
     \end{center}
     \caption{{Lightcurves for the source J0510+1800 at 8.4 GHz (top) and 4.9 GHz (middle), with their corresponding structure functions (bottom left, where the solid curve and dashed curve represent the model fits at 8.4 GHz and 4.9 GHz respectively, the dash-dot line represents $D_{noise}$ at 4.9 GHz and the dotted line represents $D_{noise}$ at 8.4 GHz) and cross-covariance function (bottom right).} \label{05161800}}
     \end{figure*}

\begin{figure*}[!htp]
     \begin{center}
     \includegraphics[scale=0.6]{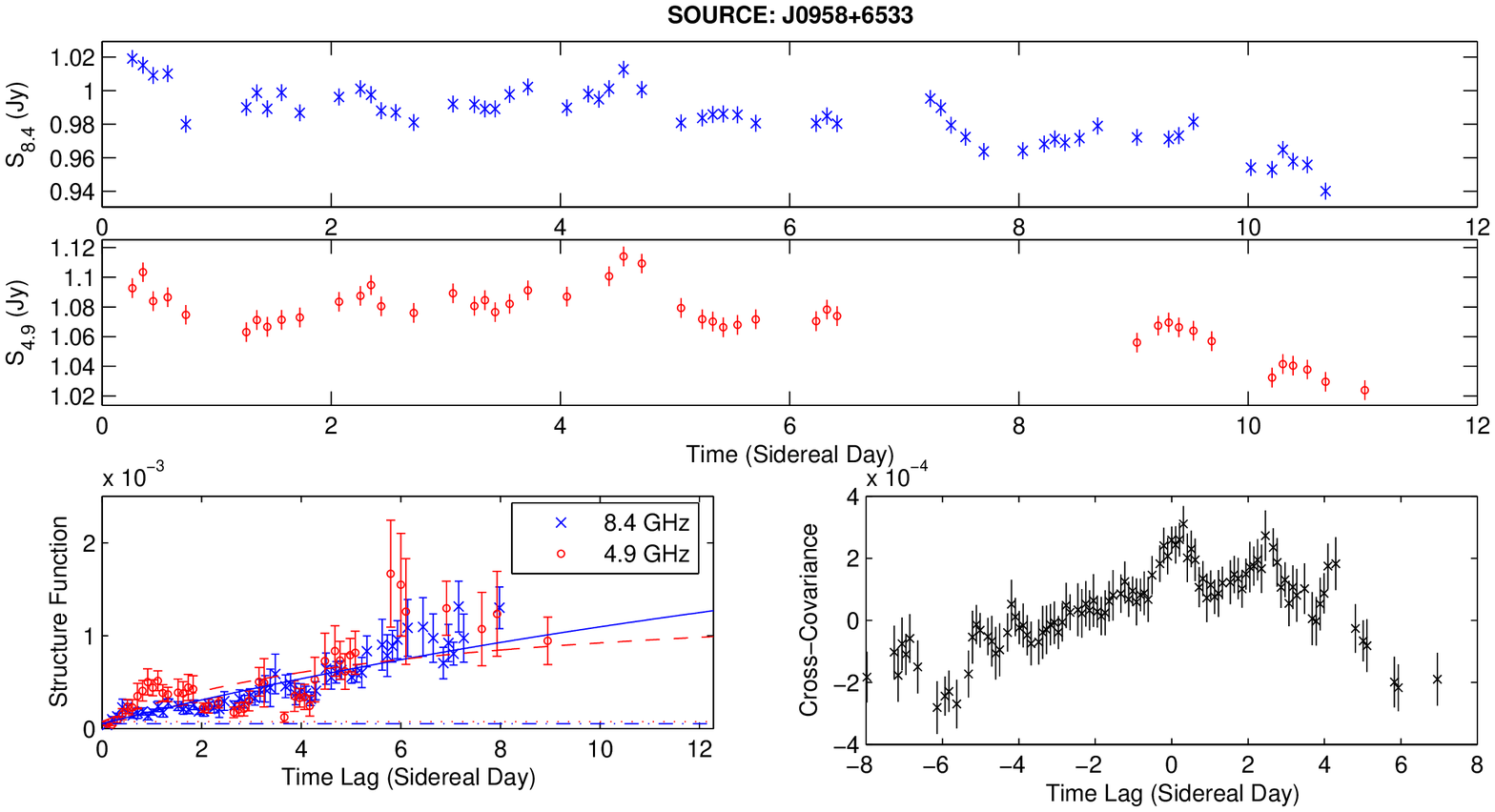}
     \end{center}
     \caption{{Lightcurves for the source J0958+6533 at 8.4 GHz (top) and 4.9 GHz (middle), with their corresponding structure functions (bottom left, where the solid curve and dashed curve represent the model fits at 8.4 GHz and 4.9 GHz respectively, the dash-dot line represents $D_{noise}$ at 4.9 GHz and the dotted line represents $D_{noise}$ at 8.4 GHz) and cross-covariance function (bottom right).} \label{09966555}}
     \end{figure*}

\begin{figure*}[!htp]
     \begin{center}
     \includegraphics[scale=0.6]{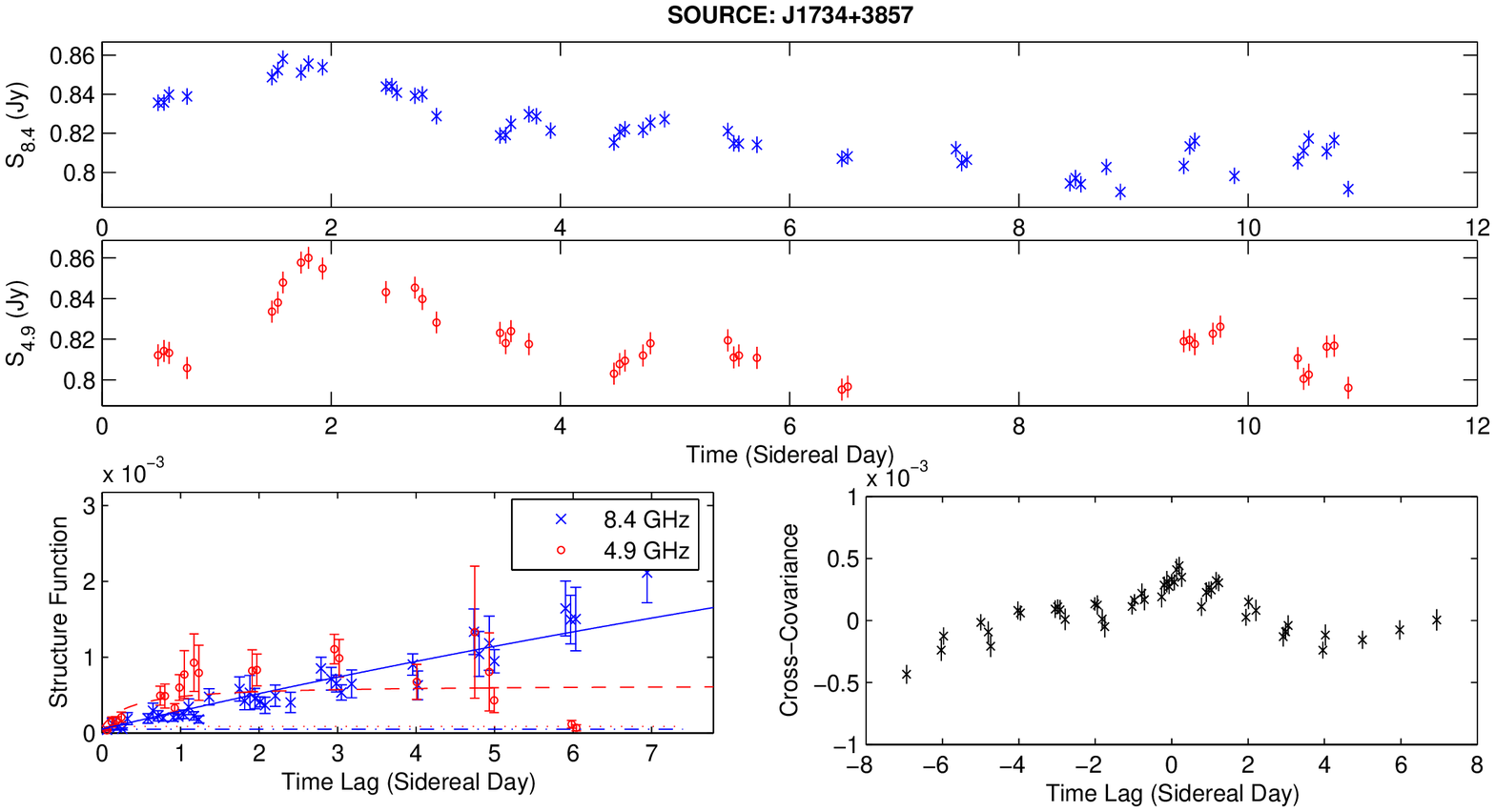}
     \end{center}
     \caption{{Lightcurves for the source J1734+3857 at 8.4 GHz (top) and 4.9 GHz (middle), with their corresponding structure functions (bottom left, where the solid curve and dashed curve represent the model fits at 8.4 GHz and 4.9 GHz respectively, the dash-dot line represents $D_{noise}$ at 4.9 GHz and the dotted line represents $D_{noise}$ at 8.4 GHz) and cross-covariance function (bottom right).} \label{17563895}}
     \end{figure*}

Figures~\ref{11982923} to \ref{17563895} show sample lightcurves of some of the variable sources. Fast scintillators such as J1159+2914 (Figure~\ref{11982923}) have variability time-scales on the order of hours. On the other hand, J0510+1800 (Figure~\ref{05161800}) is a slow variable with longer characteristic time-scales of half a day at both frequencies. Some sources have variability at multiple time-scales, where shorter and smaller amplitude variations are superposed on top of longer time-scale variations of larger amplitude. J0958+6533 (Figure~\ref{09966555}) and J1734+3857 (Figure~\ref{17563895}) are examples of such sources. This can be a result of different components in the source scintillating at different time-scales (with larger, more extended components causing slower variations and more compact components causing the faster variations). It can also be caused by a combination of short time-scale scintillation combined with longer time-scale intrinsic variability, although our analysis shows that this is not a dominant effect in our sample of sources (see Section~\ref{halpha}.

As in the analyses of the original MASIV data, the structure function (SF) is used to quantify the variability of each source. This has the advantage of being insensitive to gaps in the sampling of data (as are present in the observations analysed here), as opposed to a power spectrum analysis. Also, the SF is not as sensitive to biases resulting from errors in the estimation of the mean flux density of the source as the auto-correlation function. The observed SF at a given timelag $\tau$ is given by:
\begin{equation}\label{structurefunction}
D_{obs}(\tau)=\dfrac{1}{N_{\tau}}\sum_{j,k}[S(t_{j})-S(t_{k}-\tau)]^{2},
\end{equation}
where $S(t)$ is the measured flux density at time $t$, normalized by the mean flux density. $D_{obs}(\tau)$ is therefore a dimensionless quantity. $N_{\tau}$ is the number of pairs of flux densities with a time-lag $\tau$, binned to the nearest integer multiple of the smallest time-lag between data samples (typically 2 hours) for each source. Bins were selected for plotting the SF only if $N_{\tau}$ exceeded 20\% of the total number of sample points in the lightcurve.

Errors in the SF amplitudes at each time-lag were calculated as a standard error in the mean, given by the standard deviation of the $[S(t_{j})-S(t_{k}-\tau)]^{2}$ terms in that time-lag bin divided by $\sqrt{N_{\tau} - 1}$. We note that this method does not account for statistical errors resulting from the finite sampling of a random process, due to the limited timespan of the observations. Such statistical errors are dependent on the characteristic time-scale of the variations relative to the total observing span, increasing for sources with longer variability time-scales. A second method of calculating the SF errors was also tested, based on that used by \citet{youetal07}. In this case, the errors are given by $\sigma_{D}(\tau) = \langle D_{obs}(\tau) \rangle (\tau/{\tau_{tot}})^{1/3}$, where $\tau_{tot}$ is the total observation span, in this case 7 or 11 days, depending on which set of data was used. This estimation incorporates the fact that the number of possible pairs of flux densities that can be formed to calculate the SF generally decreases with increasing time-lag. However, it was found that errors could be underestimated for bins at small time-lags, yet have a low number of flux density pairs. Therefore, we selected the first method over the second method. 

As mentioned previously, sources in which the SF amplitudes drop at integer multiples of time-lags of a sidereal day provide a means of detecting variability patterns that are repeated daily. The SFs were therefore examined together with the lightcurves to weed out such sources for further analysis to determine the causes of these patterns.

We have also used the modulation index, $m$, to quantify the variability amplitude of the sources in various portions of the text, defined as the ratio of the rms flux density to the mean flux density of the source. The raw modulation indices (where any variability due to systematic errors have not been corrected for) of each source at both frequencies are presented in Appendix A, calculated using data from the entire observing span. Since we use the modulation index and the SF amplitude interchangably, we state upfront that the SF amplitude provides a measure of the variance, while the modulation index provides a measure of the standard deviation. The saturated SF amplitude can then be approximated as $2m^{2}$.

      \subsection{Error Estimation}\label{errors}
Several instrumental and systematic effects can contribute to the perceived variability of a source. Variability caused by such errors contribute a constant additive noise floor, $D_{noise}$, to the SF of each source. Correcting for these errors thus requires $D_{noise}$ to be subtracted from the SFs across all time-lags so that only genuine variability is retained. This is based on the assumption that the errors are independent of time-lag i.e. the errors are white. As noted in \citet{lovelletal08}, there is a possibility that some systematic errors may result in non-white errors which are dependent on time-lag. We developed and compared 3 different techniques for estimating the errors quantitatively, of which the third method (Method C), was chosen for use in the final analysis.

\subsubsection{Method A: $D_{obs}(\rm 2hr)$ as an Error Estimate}\label{errorD2hr}

A simple way of estimating $D_{noise}$ is to assume that all variability at time-scales less than 2 hours (the typical minimum time-lag between data points on the lightcurve) is not true variability by directly using $D_{noise}=D_{obs}(\rm 2hr)$ for each source, where $D_{obs}(\rm 2hr)$ is the single sample estimate of the SF at 2-hour time-lags. However, using $D_{noise}=D_{obs}(\rm 2hr)$ can lead to an overestimation of errors in some sources that do scintillate at time-scales of less than 2 hours, e.g. J1159+2914 (Figure~\ref{11982923}). On the other hand, calibration errors such as the daily repeating patterns observed in some of the sources may be underestimated, since these sources do exhibit instrument-related variability up to time-scales of a day. This method was therefore not used. 

\subsubsection{Method B: Model Fitting for the Estimation of Flux-Dependent and Flux-Independent Errors}\label{errormodel}

In the original MASIV survey, the errors were calculated based on the quadratic sum of two error components, given by the following equation \citep{lovelletal08}:
\begin{equation}\label{errorsp}
\sigma_{err,s,p}=\sqrt{(s/\bar{S})^{2}+p^{2}},
\end{equation}  
where $\sigma_{err,s,p}$ is the rms error in each flux density estimate normalized by the mean flux density of the entire length of observations, $\bar{S}$. The two error components are denoted by $s$ and $p$; $s$, which is in units of Jy, accounts for errors that are independent of the flux density of the source, including additive system noise and confusion effects, and affects mainly the weak sources; $p$, on the other hand, represents errors which are flux density dependent, such as errors in the calibration of the source as a result of residual pointing offsets, system gain variations and atmospheric absorption - these errors arise in part because there is a finite angular distance (as well as finite time interval between observations) between the target source and its calibrator. While a linear vector interpolation algorithm is used during the calibration process in AIPS to account for such effects, some residual errors will remain. Low-level variations in the calibrators themselves may also contribute to $p$. Since these errors are dependent on the source flux density, they are the dominant sources of error in the strong sources. The probability distribution of these additional variations can be assumed to be a convolution of the probability distribution of the flux density dependent errors with the distribution of the flux density independent errors, and thus can be estimated as a quadratic sum of the $s$ and $p$ error components.

\begin{figure*}[!htp]
     \begin{center}
     \includegraphics[scale=0.85]{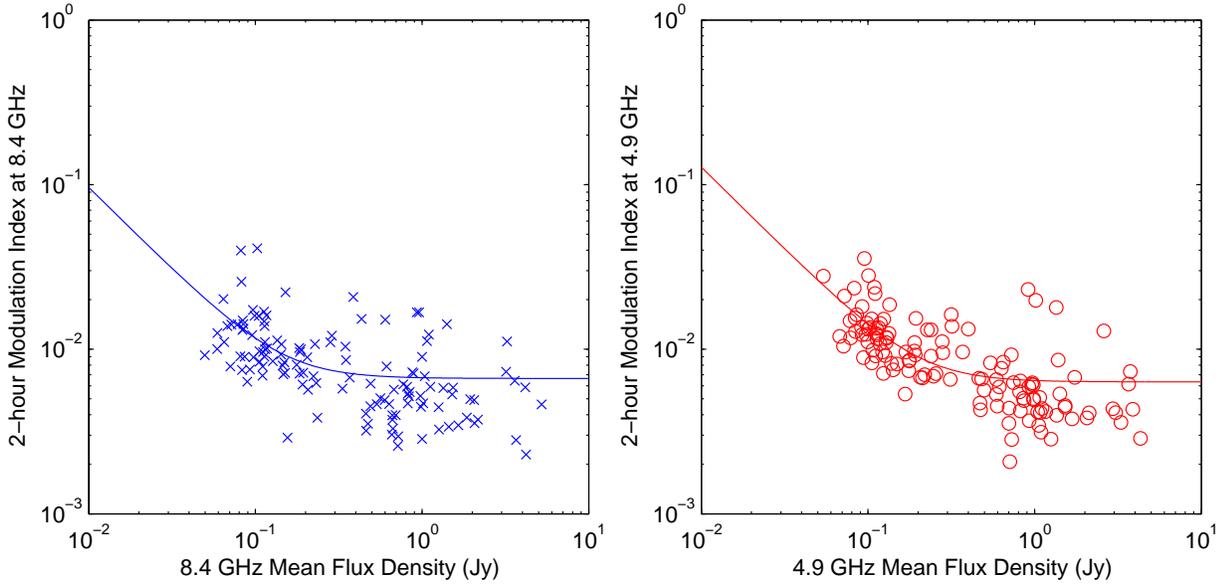}
     \end{center}
     \caption{{2-hour modulation index calculated for all sources at 8.4 GHz (left) and 4.9 GHz (right) plotted against their respective mean flux densities. The solid line represents a curve fit using Equation~\ref{errorsp}} \label{2hrmodind}}
     \end{figure*}

The values of $s$ and $p$ can be estimated by again making use of the variability of sources at its shortest measured timelag, 2 hours. The variability of each source at 2-hour time-lags is plotted against its mean flux density, as shown in Figure~\ref{2hrmodind}. In this case, the variability is quantified by the 2-hour modulation index, $m_{\rm 2hr}$, calculated from $D_{obs}(\rm 2hr)$ using $m_{\rm 2hr}=\sqrt{D_{obs}(\rm 2hr)/2}$. The equation for $\sigma_{err,s,p}$ is then used as a model fit for the resulting scatter plot (shown as a solid line), letting $s$ and $p$ be free parameters. This allows $p$ to be estimated based on the average 2-hour variability of the strong sources, and $s$ to be estimated based on the average 2-hour variability of the weak sources. Based on the curve fits, the values obtained are $s = 0.0009$ Jy and $p = 0.0068$ at 8.4 GHz, and $s = 0.0012$ Jy and $p = 0.0065$ at 4.9 GHz.  The value of $s$ obtained this way for the 4.9 GHz data is close to the value of 0.0013 Jy used in the original MASIV data, but at 8.4 GHz is lower. The reduced system noise at 8.4 GHz is to be expected given that 11 antennas (originally 13, but 2 were removed) were used in these observations as compared to the previous MASIV observations in which the VLA was sub-divided into 5 sub-arrays each with 5 or 6 antennas. However, while having a similar increase in the number of antennas, the system noise at 4.9 GHz is comparable to that in the MASIV survey due to its use of only a single IF channel. The values of $p$ used here are in the range of the values found in MASIV. These values of $s$ and $p$ are then used to calculate $\sigma_{err,s,p}$ for each source at both frequencies, from which $D_{noise}=2{\sigma_{err,s,p}}^2$ can then be subtracted from the SFs of each source.

In Method A, $D_{noise}$ is equivalent to $D_{obs}(\rm 2hr)$ for each source, but in this second method using Equation~\ref{errorsp}, about half of the sources have $D_{obs}({\rm 2hr}) > D_{noise}$, while the other half of the sources have $D_{obs}({\rm 2hr}) < D_{noise}$. Therefore, this second method of estimating $D_{noise}$ allows for about half of the sources to have real variability at time-scales less than 2 hours. While this is an improvement over the first method, it assumes that all sources have the same values of $s$ and $p$, which is definitely not the case. It also does not correct for possible low-level variations of the calibrator in an explicit manner.

\subsubsection{Method C: Source and Calibrator Dependent Error Estimates}\label{errorcalib}

This third method makes use of Equation~\ref{errorsp} as well, but uses a different approach in the calculation of the values of $s$ and $p$.

Since the amount of data flagged for each source varies and the additive errors increase as more data are flagged, it was decided that the value of $s$ would be obtained separately for each source, calculated from the standard error in the mean from the 1 minute averaging of the 3.3 second visibilities over all baselines. Since each point on the lightcurve of each source has different errors, $s$ for each source is then the average of the errors for all the points in its lightcurve. This gave values of $s$ in the range of 0.0004 to 0.0041 Jy (with a mean of 0.0007 Jy) at 8.4 GHz and 0.0006 to 0.0031 Jy (with a mean of 0.0009 Jy) at 4.9 GHz. The range of values of $s$ obtained here at both frequencies confirm that using a single $s$ value for all sources will tend to overestimate additive system noise errors in some sources while underestimate it in others.

As alluded to in Section~\ref{observations}, the fact that the daily repeating patterns were found to appear only in the stronger sources with very low variability (with $D_{obs}(\tau) < 3 \times 10^{-4}$, or raw modulation indices of $m < 1\%$), provided a clue that these errors were linked to the calibration process. Upon further examination, it was found that for sources where the SF amplitudes are greater than $3 \times 10^{-4}$ at one frequency and less than $3 \times 10^{-4}$ at the other frequency, the daily repeating patterns are observable only at the frequency with $D_{obs}(\tau) < 3 \times 10^{-4}$. Where such daily repeating patterns are superposed on top of larger, longer time-scale variations, the SF amplitudes may be much greater at longer time-lags, but between 2 hours and 1 day, the SF amplitudes are generally $ < 3 \times 10^{-4}$. The variability of these daily repeating patterns are therefore comparatively small. This led to the conclusion that these repetitive patterns were calibration errors due to pointing errors and residual gain errors from the interpolation of gain solutions between target sources and their calibrators. Though these patterns can be detected by eye when they dominate the source variability, these effects should also add to the variability of the sources dominated by real scintillation and will thus need to be corrected for.

Recognizing that the values of $p$ are calibrator dependent (due to the underlying low-level variations in the calibrator), and that the residual calibration errors need to be accounted for regardless of whether they are detectable as daily repeating patterns or not, it was decided that the value of $p$ for each source would be calculated based on the calibrator that was applied to it. To achieve this, each one of the 23 calibrators was used as a calibrator for a subset containing $N_c$ number of other calibrators with similar Local Sidereal Time (LST) coverage (with time interval between observations generally not exceeding 2 hours). $N_c$ varies for each subset and there are 23 overlapping subsets paired with 23 calibrators. After calibration, the modulation indices of all $N_c$ calibrators (we refer to them here as `target calibrators') in each of the 23 subsets were then calculated and averaged to obtain 23 values of the mean modulation index, $m_c$. Each of the 23 values of $m_c$ include both the variability of the chosen calibrator for that subset \emph{and} the variability of the other `target calibrators' in that subset. Since $m_c$ is a convolution of the probability distribution function of the flux density variations of the chosen calibrator (with a modulation index given by $m_i$) with the distribution function of the variations of the other $N_c$ `target calibrators' (with a mean modulation index given by $\langle m_{tc}\rangle$), $m_c$ is thus given by:
\begin{equation}\label{mc}
m_c = \sqrt{{m_i}^{2}+{\langle m_{tc}\rangle}^{2}}.
\end{equation}  
If we assume that the variability amplitudes of all the calibrators are roughly similar, then $m_c \approx \sqrt{2{m_i}^2}$, so that the modulation index of the chosen calibrator for each subset can be obtained as $m_i = m_c/\sqrt{2}$. Therefore, the 23 values of $m_c$ after being reduced by a factor of $\sqrt{2}$ are representative of the variability of the 23 chosen calibrators, which are then used as $p$ for all the target sources that have been calibrated by the same calibrator. We therefore have 23 sets of $p$ values distributed among the 140 target sources, depending on which calibrator was applied to them, with values ranging from 0.0048 to 0.0057 (with a mean of 0.0051) at 8.4 GHz and 0.0053 to 0.0069 (with a mean of 0.0062) at 4.9 GHz.

Another advantage of this method is that any apparent variability due to residual system gain and pointing errors are also incorporated into $p$, since these `target calibrators' were calibrated in the same manner as the actual target sources. However, since there is a larger angular distance from the chosen calibrator to most of the $N_c$ `target calibrators' as compared to the angular distances to the target sources associated with it, such residual calibration errors arising from the interpolation of the gain solutions between calibrator and target source are slightly overestimated, increasing the apparent value of $p$. A more accurate calculation would involve reducing the mean modulation indices further by a factor that accounts for the overestimated residual calibration errors, but this factor is difficult to parameterize. Further analyses with H$\alpha$, spectral index and redshift data (presented in Sections~\ref{halphasec}, \ref{spectralindexsec} and \ref{redshiftsec}) using the various estimates of $D_{noise}$ also demonstrated that any further efforts to improve the accuracy of $D_{noise}$ are unlikely to lead to further improvements in the final results for the purposes of this study.

\begin{figure}[!htp]
     \begin{center}
     \includegraphics[scale=0.5]{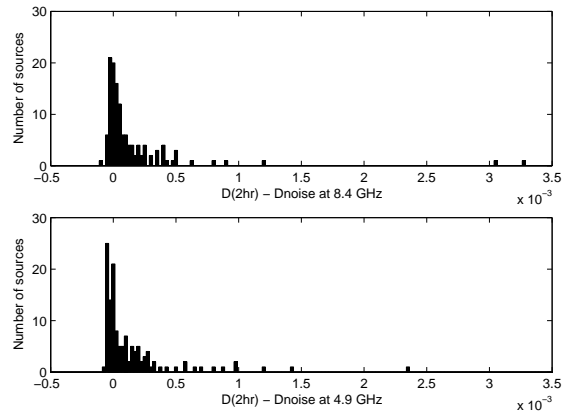}
     \end{center}
     \caption{Histogram of $D_{obs}({\rm 2hr}) - D_{noise}$, where $D_{noise}$ is estimated via Method C in Section~\ref{errorcalib}}\label{histoD2hr-Dnoise}
     \end{figure}

\begin{figure}[!htp]
     \begin{center}
     \includegraphics[scale=0.6]{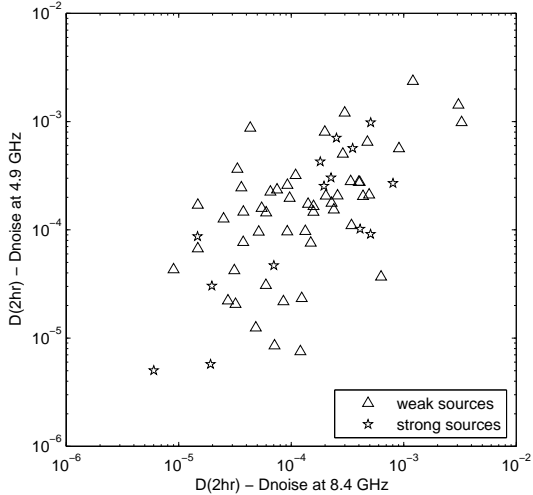}
     \end{center}
     \caption{{Scatter plot showing the correlation between $D_{obs}({\rm 2hr}) - D_{noise}$ at 4.9 GHz and $D_{obs}({\rm 2hr}) - D_{noise}$ at 8.4 GHz (for sources where the quantities are positive at both frequencies) based on the error estimation in Method C (Section~\ref{errorcalib}).}\label{d2hr-dnoise2freq}}
     \end{figure}

Plotting the histograms of $D_{obs}({\rm 2hr}) - D_{noise}$ (Figure~\ref{histoD2hr-Dnoise}) shows distributions with peaks located close to 0 at both frequencies. For a sample of non-variable sources, one would expect a Gaussian distribution with a peak at 0. Our plots show a tail towards the right of the plot, caused by the presence of variable sources in the sample. To confirm this, we plotted $D_{obs}({\rm 2hr}) - D_{noise}$ at 4.9 GHz against $D_{obs}({\rm 2hr}) - D_{noise}$ at 8.4 GHz (Figure~\ref{d2hr-dnoise2freq}), which demonstrates a clear correlation between these two quantities. We obtained a statistically significant (at a significance level of 0.05) Pearson's linear correlation coefficient of 0.63, with a p-value of $5.5 \times 10^{-8}$. As the 8.4 GHz and 4.9 GHz subarrays comprise different antennas and therefore have different uv-coverages, it is unlikely that this correlation is due to antenna-based, array-based, or sky-based errors (ie. low level confusion). The errors would therefore have been overestimated for the sources with $D_{obs}({\rm 2hr}) - D_{noise} > 0$ had Method A been used. 

\begin{figure*}[!htp]
     \begin{center}
     \includegraphics[scale=0.85]{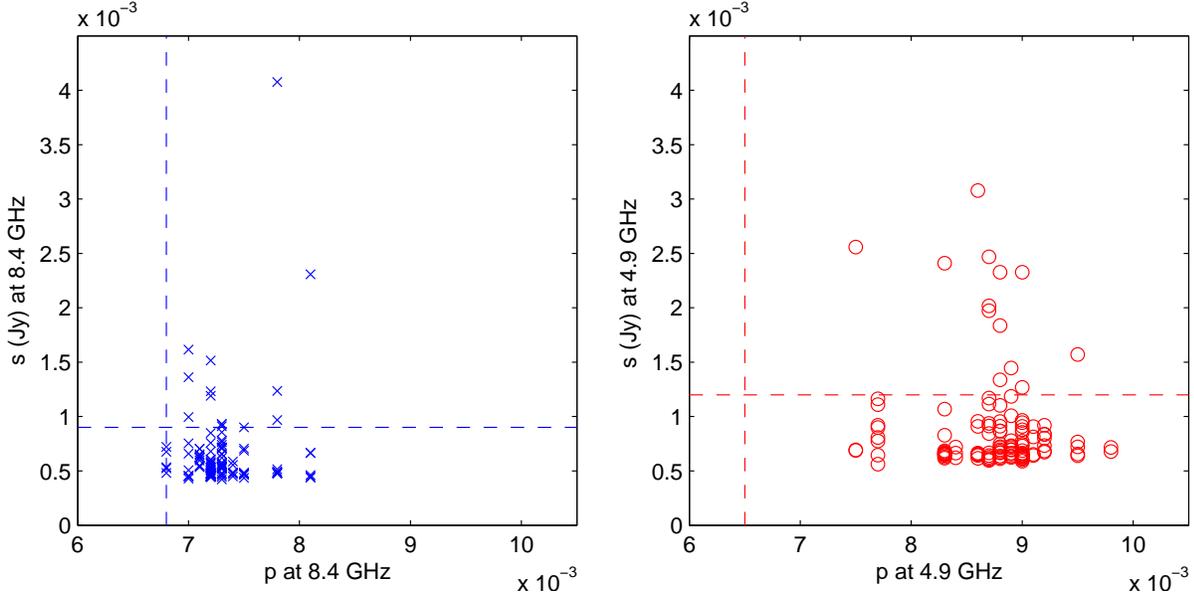}
     \end{center}
     \caption{{Scatter plot of $s$ against $p$ at 8.4 GHz (left) and 4.9 GHz (right) based on the error estimation in Method C (Section~\ref{errorcalib}). The dashed lines represent estimated values of $s$ and $p$ from Method B (Section~\ref{errormodel})}\label{svsp}}
     \end{figure*}

As a further comparison between this method with Method B, Figure~\ref{svsp} shows the scatter plot of $s$ vs $p$ estimated through Method C, with the dashed lines showing the values of $s$ and $p$ estimated via Method B. As expected, with the exception of a few outliers, $s$ is lower at 8.4 GHz than at 4.9 GHz. It appears that $s$ is generally overestimated when Method B is used. This is most likely due to the fact that $s$ in Method B is determined by the 2-hour modulation indices of the weak sources, which are known to scintillate more than the strong sources \citep{lovelletal08}. On the other hand, $p$ is clearly underestimated in Method B as it does not account for low-level calibrator variability and residual calibration errors that have variability time-scales longer than 2 hours. $p$ appears to be larger at 4.9 GHz than at 8.4 GHz, whereas one would expect residual pointing errors and antenna-based gain related errors to be generally smaller at longer wavelengths. This can be explained by the removal of data from 2 antennas in the 8.4 GHz subarray in which the pointing errors appeared the worst, as mentioned briefly in Section~\ref{observations}. The removal of these antennas resulted in a negligible increase in $s$. We also attempted to remove data from 2 antennas in the 4.9 GHz subarray in an attempt to reduce $p$, but resulted in a similar magnitude increase in $s$ (recall that the 4.9 GHz observations were conducted at half the bandwidth of the 8.4 GHz observations). We therefore retained all antennas in the 4.9 GHz subarray.

\begin{figure*}[!htp]
     \begin{center}
     \includegraphics[scale=0.7]{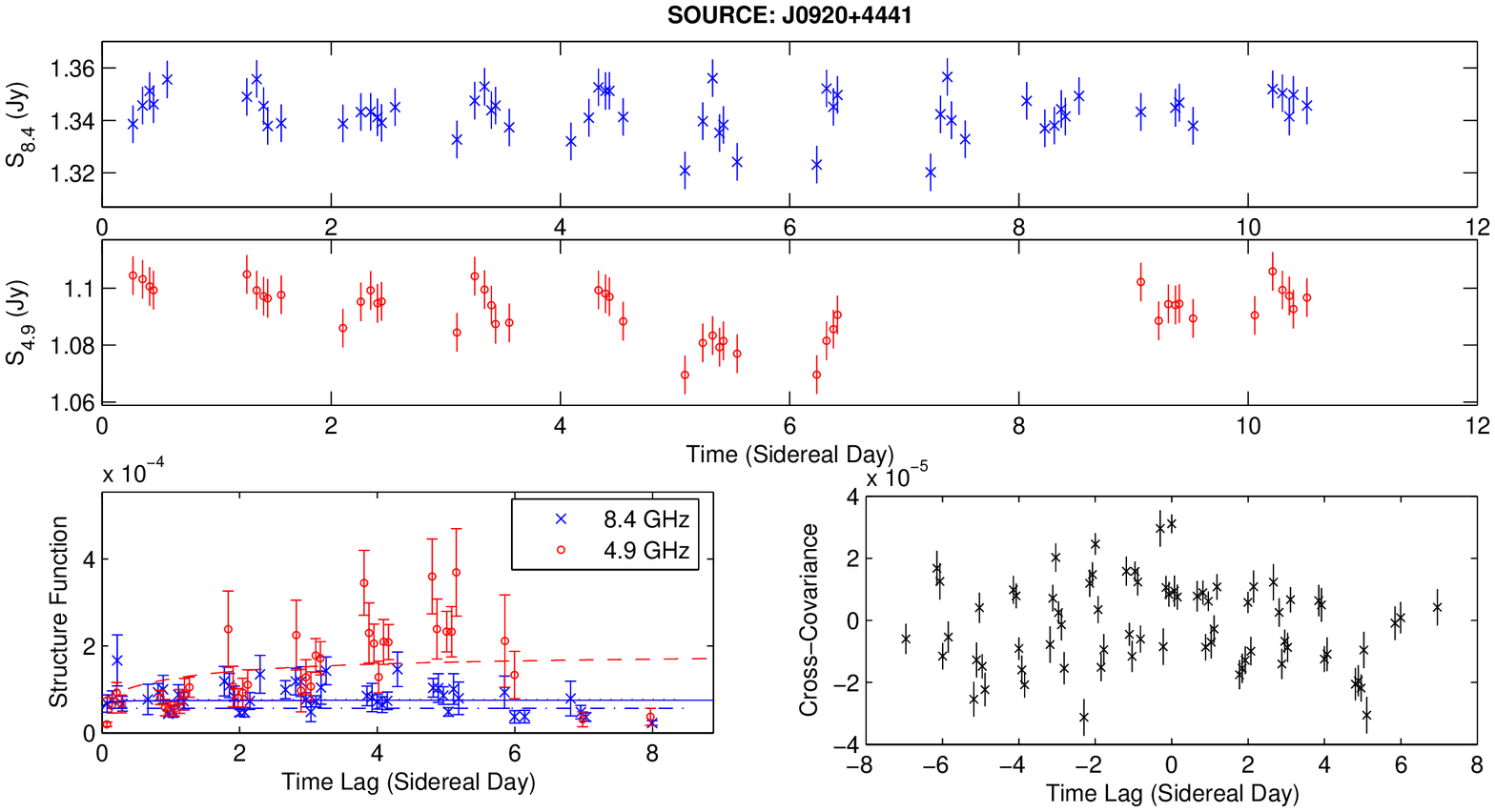}
     \end{center}
     \caption{{8.4 GHz (top) and 4.9 GHz (middle) lightcurves, structure functions (bottom left, where the solid curve and dashed curve represent the model fits at 8.4 GHz and 4.9 GHz respectively, the dash-dot line represents $D_{noise}$ at 4.9 GHz and the dotted line represents $D_{noise}$ at 8.4 GHz) and cross-covariance function (bottom right) for the source J0920+4441, as a demonstration of the error estimation and correction used.} \label{09334468errors}}
     \end{figure*}

Figure~\ref{09334468errors} demonstrates the effectiveness of the error estimation and correction described in Method C; it shows a source with very low variability. Daily repeating patterns are observed at both frequencies, particularly between 2 to 6 sidereal days. Their effect on the SF is modelled successfully by $D_{noise}$ as can be seen in the corresponding SF plots. At 8.4 GHz, $D_{obs}(\tau)$ is distributed around $D_{noise}$ (shown as a dash-dot line), while for 4.9 GHz, $D_{obs}(\tau)$ is close to $D_{noise}$ (shown as a dotted line) for time-lags up to about 3 days before rising to double the value of $D_{noise}$. Although the SF amplitudes at 4.9 GHz are not high, the daily repeating patterns are superposed on top of longer term variations, which are not visible in the 8.4 GHz lightcurves.

Finally, a total of 11 sources were eventually removed from our sample; J1535+6953 had a very low mean flux density ($\approx 30$ mJy) in the current 2009 epoch, and upon further investigation, we found that its mean flux density had been steadily decreasing from 75 mJy in the first MASIV epoch to 60 mJy in the fourth MASIV epoch; the other 10 sources were found to have daily repeating patterns that varied with SF amplitudes exceeding $3 \times 10^{-4}$, possibly due to real confusion and resolution effects that were not detectable in the snapshot images and uv-data. In the latter group, their errors were not well-characterized by the method of error estimation described above, and could not be removed by any other means.

      \subsection{Structure Function Fitting}\label{sffit}
Assuming that variability due to ISS approaches a stationary stochastic process when observed over a duration much longer than its characteristic time-scale, it is expected that the true SF, $D(\tau)$, will increase with time-lag and saturate at twice the true variance. Therefore, a simple model can be used to fit the SF data, given by:
\begin{equation}\label{sfmodel}
D_{mod}(\tau)=2m^{2}\dfrac{\tau}{\tau+{\tau}_{char}}+D_{noise},
\end{equation}
where $2m^{2}$ is the value at which the SF saturates, and is related to the modulation index, $m$, of the source.  $\tau_{char}$, is the characteristic time-scale at which the SF reaches half of its value at saturation. $2m^{2}$ and $\tau_{char}$ are both free parameters of the model. This is similar to the model used in the original MASIV data analyses, and assumes that the ISS is caused by a turbulent ISM distributed uniformly through a thick scattering region. Further details on the justification for its use can be found in Appendix A of \citet{lovelletal08}. The model also assumes that variations due to systematic and instrumental errors contribute an additive term, $D_{noise}$, to the overall variability. $D_{noise}$ is determined for each source using the method explained in the previous subsection. The true SF, $D(\tau)$ can thus be obtained by subtracting $D_{noise}$ from the model SF, $D_{mod}(\tau)$.

In fitting the model, each $D_{obs}(\tau)$ is weighted by $\sqrt{\langle D_{obs}(\tau) \rangle /\sigma_{D}(\tau)}$, where $\sigma_{D}(\tau)$ is the error of the SF estimate at that particular time-lag. The result is that these values of $D_{obs}(\tau)$ with smaller errors will have larger weights. If the estimation of the errors were accurate, the weights should be proportional to $1/\sigma_{D}(\tau)$. Due to uncertainties in the estimation of the SF errors, $1/\sqrt{\sigma_{D}(\tau)}$ is used instead. The weights are further normalized by $ \langle D_{obs}(\tau) \rangle$, which is the ensemble average of all the SF estimates for the source at all timelag bins whose number of pairs of flux densities are above the threshold value (see Section~\ref{sf}). The fit was carried out using a non-linear least squares method.

Sample SFs are shown together with their corresponding model fits (represented by the solid curve and dashed curve for 8.4 GHz and 4.9 GHz respectively) in the bottom left corners of Figures~\ref{11982923} to \ref{17563895}. The dash-dot and dotted lines represent the $D_{noise}$ values for 8.4 GHz and 4.9 GHz respectively. The SF for a fast scintillator such as J1159+2914 (Figure~\ref{11982923}) reaches saturation on a time-scale of a couple of hours. Some sources such as J0958+6533 (Figure~\ref{09966555}), however, have yet to saturate even at time-lags of 8 days. Some SFs have a periodic trend, which is caused by the lightcurve having a periodic structure (as can be seen for J0510+1800 in Figure~\ref{05161800}) within the limited timespan of the observations. If the timespan of the observations were to be increased, the fluctuations would become randomized and $D_{obs}(\tau)$ should approach that of $D_{mod}(\tau)$, demonstrating the deficiencies in the estimation of the error bars.

For the purpose of statistical analyses in the following subsections, unless otherwise stated, the SF amplitudes at 4 days were used, obtained from the model fit with $D_{noise}$ subtracted, given by $D({\rm 4day}) = D_{mod}({\rm 4day}) - D_{noise}$. Instead of using Equation~\ref{sfmodel}, we have used an equivalent functional form:
\begin{equation}\label{sfmodel2}
D_{mod}(\tau)=D({\rm 4day})\dfrac{1+{\tau}_{char}/4}{1+{\tau}_{char}/\tau}+D_{noise},
\end{equation}
so that $D(\rm 4day)$ becomes a fitted parameter instead of $2m^{2}$. This way, the 95\% confidence bounds of $D(\rm 4day)$ from fitting the model can be obtained directly, which we use as an estimate of the errors in $D(\rm 4day)$. As opposed to using the single time-lag estimates ($D_{obs}(\rm 4day)$), the model fits provide better statistical representation, since they made use of the SF amplitudes at all available time-lags. The SF amplitudes at 4 days were chosen as standard characterization of source variability to ensure that a large majority of the SFs had reached saturation, and that there were still sufficient number of pairs at the nearby bins to provide reliable SF fits. While choosing $D(\rm 10day)$ or $D(\rm 11day)$ as standard characterization of the variability will maximize the number of sources with saturated SFs, the fitted curve may not be as reliable at those time-lags.

$D(\rm 4day)$ and $\tau_{char}$ at both frequencies are presented for each source in Appendix A. While the 11 day observations provide better constraints on $\tau_{char}$, its errors are still very difficult to estimate. Therefore the values of $\tau_{char}$ for a source at a particular frequency are shown only if $D(\rm 4day)$ exceeds $3 \times 10^{-4}$, as sources with barely detectable variability tend to be dominated by sytematic errors and noise so that estimates of $\tau_{char}$ may be inaccurate. We are also unable to estimate $\tau_{char}$ for sources in which the SFs do not show signs of saturating and we simply note in Appendix A that these sources have $\tau_{char} > 11$ days. 
      
      \subsection{Interpretation as ISS}\label{halphasec}
We investigate here whether the variations observed in the lightcurves and SFs are a result of ISS. Since these observations were carried out over a period of 11 days as compared to the 3 or 4 day observations in the original MASIV survey, it is also important to determine if the source variability at longer time-scales can still be attributed to ISS rather than being intrinsic variations.

\begin{figure*}[!htp]
     \begin{center}
     \includegraphics[scale=0.8]{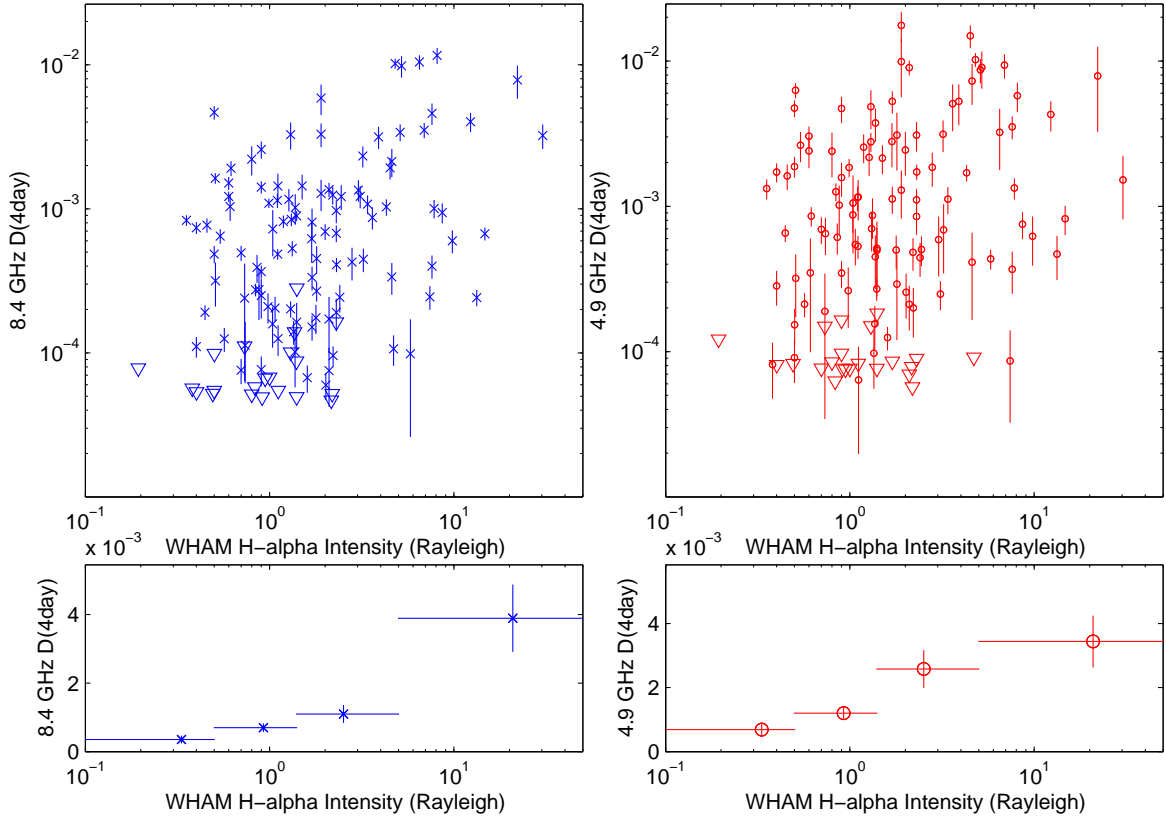}
     \end{center}
     \caption{{Scatter plot of $D({\rm 4day})$ at 8.4 GHz (top left) and 4.9 GHz (top right) plotted against WHAM H$\alpha$ intensities. The triangles represent upper limits of $D({\rm 4day})$ for sources where $D({\rm 4day}) < D_{noise}$. Corresponding binned averages of $D({\rm 4day})$ are also shown for 8.4 GHz (bottom left) and 4.9 GHz (bottom right).} \label{halpha}}
     \end{figure*}

To determine this, $D({\rm 4day})$ for each source is plotted against its line of sight H$\alpha$ intensity in Rayleighs, obtained from the corresponding $1^{\circ}$ grid in the Wisconsin H$\alpha$ Mapper (WHAM) Northern Sky Survey database \citep{haffneretal03}. The H$\alpha$ intensities provide an estimate of the emission measure of the ionized ISM in the direction of the source. The scatter plots obtained are shown in Figure~\ref{halpha}. For sources where $D({\rm 4day})$ was found to be less than $D_{noise}$, we have used $D_{noise}$ as an upper limit of the variability amplitude of the source (denoted as triangles in the scatter plots). The bottom portion of Figure~\ref{halpha} plots the average $D({\rm 4day})$ in 4 separate bins. The correlation between $D({\rm 4day})$ and H$\alpha$ intensity for both frequencies can be seen. While the plots shown here made use of the data from the entire 11 day duration of the observations, the correlation holds true even when only data from the first 6 days were used. The non-parametric Kendall's tau test confirms positive correlations between $D({\rm 4day})$ and H$\alpha$ intensities at both frequencies, with rank correlation coefficients of 0.23 at 8.4 GHz and 0.18 at 4.9 GHz. Although the correlations are weak, they are statistically significant, with p-values of $1.2 \times 10^{-4}$ and $3.0 \times 10^{-3}$ at 8.4 GHz and 4.9 GHz respectively. Here and in all subsequent analyses, we have chosen the standard significance level of 0.05.  

As a further test, the single sample estimates of the observed SF, $D_{obs}(\tau)$, with $D_{noise}$ subtracted and $\tau$ = 1, 2, 3 ... 7 days, were each used in succession to plot against the WHAM H$\alpha$ intensities. The significant correlation of the SFs with H$\alpha$ intensity is retained for all time-lags when data from all 11 days were used. Similar results were obtained for $\tau$ = 1,2,3 and 4 days when data from only the first 6 days of observations were used. It can thus be reasonably concluded that the observed flux density variations in this study, including those at longer time-scales of up to 7 days, are predominantly linked to ISS.

      \subsection{Correlation of ISS Across Frequencies}\label{freqcorr}
\begin{figure}[!htp]
     \begin{center}
     \includegraphics[scale=0.7]{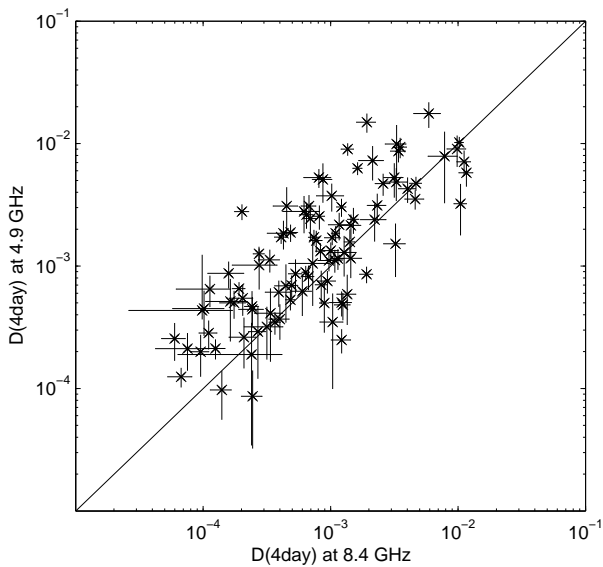}
     \end{center}
     \caption{{4-day Structure Functions at 4.9 GHz, $D_{4.9}({\rm 4day})$, vs 4-day Structure Functions at 8.4 GHz, $D_{8.4}({\rm 4day})$. The solid line represents $D_{4.9}({\rm 4day}) = D_{8.4}({\rm 4day})$ as a demonstration of the correlation of the SFs at both frequencies and that $D_{4.9}({\rm 4day})$ is generally larger than $D_{8.4}({\rm 4day})$.} \label{SFvsSF}}
     \end{figure}

According to the theory of interstellar scintillation, weak scintillation should be correlated over a wide bandwidth, with a decorrelation bandwidth on the order of the observing frequency \citep{narayan92}. Although the 4.9 GHz observations are near the transition between weak and strong scintillation at mid-Galactic latitudes \citep{walker98, walker01}, some form of correlation is still expected to exist between the variability at 4.9 GHz and 8.4 GHz. In Figure~\ref{SFvsSF}, $D({\rm 4day})$ at 4.9 GHz ($D_{4.9}({\rm 4day})$) is plotted against $D({\rm 4day})$ at 8.4 GHz ($D_{8.4}({\rm 4day})$) on a log scale for sources with $D({\rm 4day}) > D_{noise}$ at both frequencies, showing that the source variability amplitudes are well-correlated between both frequencies. 
     
While correlation of the variability patterns between the lightcurves at both frequencies can be clearly discerned by eye for some sources, the cross-covariance function provides a more quantitative means of detecting such a correlation, obtained as:
\begin{equation}\label{crossfunction}
C(\tau)=\dfrac{1}{N_{\tau}}\sum_{j,k}[S_{4.9}(t_{j})-\mu_{S_{4.9}}][S_{8.4}(t_{k}-\tau)-\mu_{S_{8.4}}],
\end{equation}
where $S_{4.9}$ and $S_{8.4}$ are the normalized flux densities at 4.9 GHz and 8.4 GHz respectively, and $N_{\tau}$ is the number of pairs of flux densities with a time-lag of $\tau$. $\mu_{S_{4.9}}$ and $\mu_{S_{8.4}}$ are the average values of $S_{4.9}$ and $S_{8.4}$ over the entire observation span. As in the calculation of the discrete SF in Equation~\ref{structurefunction}, time-lag bins at integer multiples of the smallest time-lag between data points are used, with time-lag bins selected for plotting only if they comprise of more than 20\% of the total number of points in the lightcurve.

The cross-covariance functions between the lightcurves at both frequencies are shown together with their associated lightcurves in the bottom right corners of Figures~\ref{11982923} to \ref{17563895}. For the majority of the variable sources, the cross-covariance peaks at time-lags of $0.00 \pm 0.04$ days. About 20\% of the sources in the sample do not show any evidence of correlation in the lightcurves or have a very weak correlation (the cross-covariance function peaks at an amplitude $< 1 \times 10^{-4}$). In such cases, the variability also tends to be very weak, with very low SF values. We conclude that the variations seen in these lightcurves are most likely attributable to noise. Another possible explanation is that there is an offset in the positions of the source cores at the two frequencies, but these offsets are perpendicular to the direction of the relative motion between the interstellar scattering screen and the Earth, which weakens the correlation. The lightcurves may also be weakly correlated in cases where the variations are due to strong refractive scintillation where the variability timescales can be different at both frequencies.

The sources that show time delays in scintillation patterns at the two frequencies are of particular interest. This is discernible by a shift of the peak of the cross-covariance function towards a non-zero time-lag. Such a delay in the lightcurves between observing frequencies has been previously observed, and has been interpreted as being caused by a small shift in the position of the optically thick compact core when observed at different frequencies  \citep{bignalletal03}. Such core shifts have been observed on milliarcsecond scales in VLBI images of extragalactic radio sources at different frequencies \citep{kovalevetal08, tzioumisetal10}. A list of sources in which the cross-covariance function of the lightcurves at 4.9 and 8.4 GHz peaks at a non-zero time-lag is shown in Table~\ref{freqcorrtable}. For this list, only sources with $D_{obs}(\tau) > 3 \times 10^{-4}$ at both frequencies and whose cross-covariance peaks at time-lags greater than twice the size of the smallest time-lag bin were selected. A negative time-lag indicates that the 8.4 GHz scintillation pattern is lagging behind the 4.9 GHz scintillation pattern. 

\input{freqcorrtab.tex}

Taking the source J0510+1800 (Figure~\ref{05161800}) as an example, there appears to be a time delay, $\Delta t$ of about $0.24 \pm 0.04$ days between the 4.9 GHz and 8.4 GHz variability patterns. Assuming typical scattering screen distances of $L = 500$ pc and screen velocities of $v = 50 $ $\rm kms^{-1}$, it is estimated that there is an angular separation of approximately $14 \pm 2$ microarcseconds between the position of the cores at 4.8 and 8.4 GHz (the component parallel to the direction of screen velocity). The angular separation of the cores for the remaining sources in Table~\ref{freqcorrtable} can be calculated using the following: 
\begin{equation}\label{angularseparation}
\theta = 14 \left( \dfrac{\Delta t}{0.24\rm days} \right) \left( \dfrac{v}{50 \rm kms^{-1}} \right) \left( \dfrac{L}{500 \rm pc} \right)^{-1} \mu \rm as,
\end{equation}
where the parameters of the scattering screen are normalized by their typical values, and $\Delta t$ is obtained from the observations.

VLBI measurements of core shifts of extragalactic radio sources between frequencies of 2.3 and 8.6 GHz by \citet{kovalevetal08} have yielded angular separations ranging from 0.1 to 1.4 milliarcseconds. In an ideal scenario, observations of source scintillation at 2-hour intervals (thus providing a minimum observable time-lag of 2 hours), enable core shifts to be probed down to about 5 microarcseconds, well beyond the capabilities of current VLBI. In practice, however, such observations are hampered by the dominance of systematic errors at these small time-lags. Using such small bin sizes for the time-lags in cross-covariance function analysis leads to large statistical errors. Conversely, using larger time-lag bins reduces the time-resolution that such a technique can probe. The fact that interstellar scattering in itself leads to shifts in apparent source positions adds to the complexity of the problem.
      
      \subsection{ISS and Source Spectral Index}\label{spectralindexsec}
In the MASIV survey, the SF amplitudes showed only a very weak trend with respect to the estimated source spectral index. A significant limiting factor in the analysis was that the flux density data used in the estimation of the spectral indices --- the 1.4 GHz NVSS data \citep{condonetal98}, the 8.4 GHz JVAS data \citep{patnaiketal92, browneetal98, wilkinsonetal98} or CLASS data \citep{myersetal95} --- were non-simultaneous at different frequencies, and are thus affected by changes in source properties. Also, being highly compact and intrinsically variable, the sources could have undergone changes to their structure and other intrinsic properties in the time between the observations through which the spectral indices were derived and the 4 epochs of the MASIV survey.

\begin{figure*}[!htp]
     \begin{center}
     \includegraphics[scale=0.8]{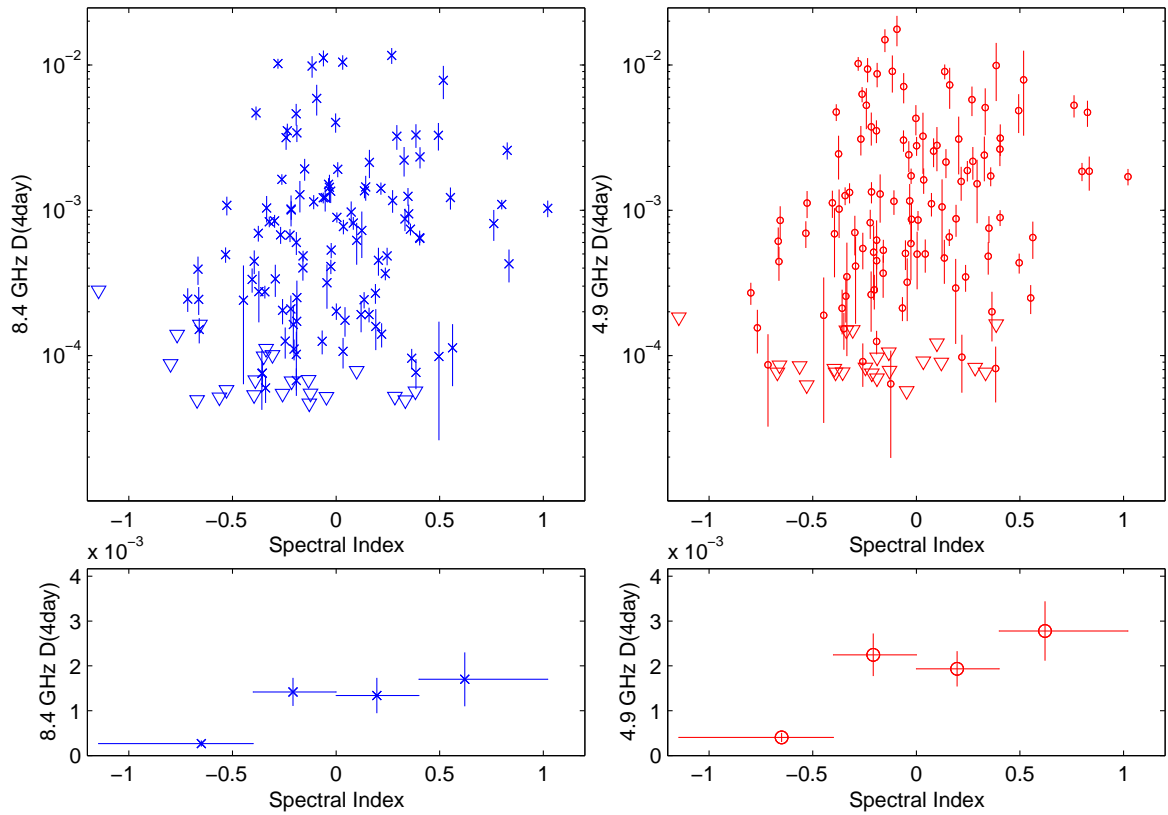}
     \end{center}
     \caption{{$D({\rm 4day})$ at 8.4 GHz (scatter plot at top left and binned plots at bottom left) and 4.9 GHz (scatter plot at top right and binned plots at bottom right) plotted against source spectral index. The triangles represent upper limits of $D({\rm 4day})$ for sources where $D({\rm 4day)} < D_{noise}$.} \label{specindex}}
     \end{figure*}

The dual frequency observations obtained in this present study enable the spectral index of each source to be estimated, bearing in mind the limitation of having the spectral indices determined by only two frequency measurements of the flux densities, which are also modulated by significant ISS. Figure~\ref{specindex} shows $D(\rm 4day)$ at both frequencies plotted against the source spectral indices calculated from the mean flux densities at both frequencies. The convention used to define the spectral index, $\alpha$, is $S \propto \nu^{\alpha}$. It is interesting to note that while only nominally `flat spectrum' sources were selected for our sample, based on the aforementioned less reliable estimations of the spectral indices, the scatter plots reveal that a few of the sources show $\alpha < -0.4$ or $\alpha > 0.4$, attesting to the variable nature of the sources. Furthermore, calculating the apparent spectral indices using each individual data point on the lightcurves of each source at both frequencies shows that the spectral indices vary even within the 11 day time-span of the observations, with a standard deviation of up to 0.13 from the mean spectral index (0.04 on average for all the sources).

While the binned plots show no clear trends for sources with $\alpha > -0.4$ at both frequencies, with only a minimal increase in the mean spectral index above $\alpha > 0.4$, there is a clear reduction of scintillation amplitudes below a spectral index of -0.4. The non-parametric Kendall's tau test gives correlation coefficients of 0.20 at 4.9 GHz and 0.19 at 8.4 GHz, with p-values of $9.0 \times 10^{-4}$ and $1.8 \times 10^{-3}$ respectively, showing that the correlations are statistically significant. However, when performing the same test using only sources with $\alpha > -0.4$, the correlation coefficient drops to 0.13 and only has a marginal statistical significance (with a p-value of 0.05). Again, these trends were also observed when data from only the first 6 days of observations were used. An explanation is that the flat-spectrum sources are dominated by very high-brightness temperature, optically thick, synchrotron self-absorbed components, thus most of their flux densities are confined to ultra-compact, microarcsecond scale cores. On the other hand, the steep spectrum sources are dominated by optically thin, compact milliarcsecond components that have lower brightness temperatures. While it is well known that scintillating sources tend to have flat or inverted spectra, and that steep spectrum sources do not scintillate \citep{heeschen84}, we note that the steep spectrum sources in our sample are unlike the 3C sources reported by \citet{heeschen84}. The steep spectrum sources in our sample do scintillate, but their flux densities are dominated by very compact milliarcsecond components as opposed to the microarcsecond components, so that their scintillation amplitudes are highly suppressed relative to that of the flat spectrum sources. Any bias in the distribution of such steep spectrum sources in the high and low redshift source samples will affect the interpretation of the redshift dependence of ISS.

      \subsection{Comments on Individual Sources}\label{individual}

We present here a discussion on the observed properties of selected sources that may be of interest to the reader. Some of these are well-known sources often targeted for VLBI and scintillation studies. They also highlight the complexity of interpreting the underlying physics in AGN scintillation. 
      
      \subsubsection{J1159+2914}

The optically violently variable quasar J1159+2914 (QSO 1156+295) (Figure~\ref{11982923}) was initially found to be scintillating at radio wavelengths (5 GHz) with 5.6\% rms variations and with a timescale of $\lesssim$ 12 hours by \citet{lovelletal03}. 15 GHz VLBI observations in 2007 \citep{savolainenkovalev08} found the source to be scintillating with a modulation index of 13\% and at a timescale of 2.7 hours (calculated as the average of the peak-to-trough and trough-to-peak time). It was uncertain as to why the later rms variations were larger at 15 GHz than ealier at 5 GHz. It was proposed that either the source was more compact during the VLBI observations than during the MASIV survey, or that the variability results from strong scintillation rather than weak scintillation. Our simultaneous dual frequency observations indicate rms variations of 3.6\% at 4.9 GHz and 1.9\% at 8.4 GHz, so it is unlikely that the source is undergoing strong scintillation at the present epoch. The estimated timescales are 2 hours at 8.4 GHz and 4 hours at 4.9 GHz. However, it is difficult to make straightforward comparisons based on the modulation indices since VLBI measures the flux density at milliarcsecond scales whereas the VLA flux includes larger scale components. The flux density of the actual scintillating component is in turn an unknown and variable fraction of the VLBI and VLA flux density. Examining the unnormalized variations give 0.6 Jy peak-to-trough variations at 5 GHz in \citet{lovelletal08}, 0.7 Jy peak-to-trough variations at 15 GHz in \citet{savolainenkovalev08}, and 0.4 Jy peak-to-trough variations in our observations at both frequencies.  
      
      \subsubsection{J1819+3845}

The well-known quasar J1819+3845 has been observed to consistently display 20 to 35\% rms variations in its flux density since its extreme variability was discovered in 1999 \citep{dennett-thorpedebruyn02,dennett-thorpedebruyn03}, with scintillation timescales down to 15 minutes \citep{macquartdebruyn07}. This rapid scintillation is attributed to the presence of a nearby scattering region about 4 to 12 pc from the Earth. Surprisingly, the scintillations appeared to have stopped abruptly when no variability was detected in a VLBI observation in 2008 \citep{cimo08}. This can result from a change in the microarcsecond structure of the source, or from a change in the nearby scattering cloud (possibly moving away). Our observations show significant rms variations of about 2\% at both frequencies, and at 8.4 GHz is superposed on top of longer timescale variations $> 11$ days. Either the source has begun scintillating again (albeit at a lower level and at a slower timescale) after the abrupt halt, or these low-level scintillations were undetectable using the technique used by \citet{cimo08}.
      
      \subsubsection{J1919+3159}

J1919+3159 exhibits the strongest variability for a source amongst our $z > 3$ sample. The long timescale variations ($> 11$ days at both frequencies) suggest a relatively large angular size. The fact that the 8.4 GHz $D(\rm 4day)$ is larger than the 4.9 GHz $D(\rm 4day)$ appears consistent with a source undergoing weak ISS with its effects suppressed further at the lower frequency due to scatter broadening in the IGM. However, the line-of-sight H$\alpha$ intensity of 6.5 Rayleighs and Galactic latitude of only 8.6 degrees indicate that the more plausible explanation is that the source is undergoing strong refractive scintillation, which would also have long timescale variations and larger variability amplitudes at the higher frequency. The slow variations can also be attributed to intrinsic effects, although the observed $\approx 1$ day lag in the 8.4 GHz lightcurve compared to that at 4.9 GHz renders this unlikely. This example demonstrates the complex physics involved in the interpretation of the data, the understanding of which will be critical in our efforts to investigate the redshift dependence of ISS. 

	  \subsubsection{J0800+4854, J1328+6221, J1549+5038 and J1931+4743}

These four sources represent some of the fastest scintillators in our sample, exhibiting 3 to 7\% rms variations at 4.9 GHz with variability timescales estimated to be on the order of $\lesssim$ 2 hours at both frequencies. The lightcurves of these sources show strong scatter in the flux densities with time, but are well correlated accross both frequencies. The very rapid, intra-hour scintillators were found to be rare from the MASIV observations (no new sources of a similar nature were discovered) \citep{lovelletal08}, interpreted as being caused by very rare, nearby scattering regions, similar to that for J1819+3845. Since our observations have a lower limit of 2 hours between each pointing, it will be of interest to conduct follow-up observations at shorter timescales to obtain better estimates of the variability timescales of these sources.
                    
      \subsection{Redshift Dependence of ISS}\label{redshiftsec}
\begin{figure*}[!htp]
     \begin{center}
     \includegraphics[scale=0.8]{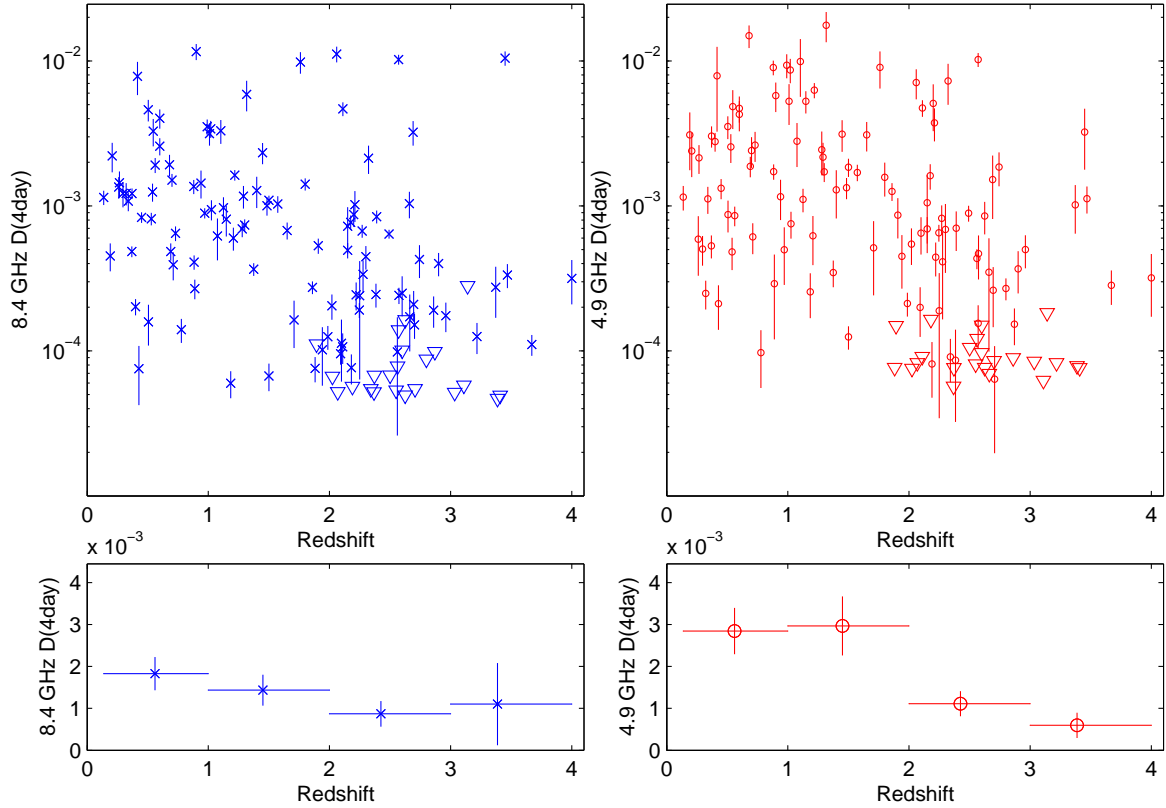}
     \end{center}
     \caption{{$D({\rm 4day})$ at 8.4 GHz (scatter plot at top left and binned plots at bottom left) and 4.9 GHz (scatter plot at top right and binned plots at bottom right) plotted against source redshift. The triangles represent upper limits of $D({\rm 4day})$ for sources where $D({\rm 4day}) < D_{noise}$.} \label{redshift}}
     \end{figure*}

\begin{figure}[!htp]
     \begin{center}
     \includegraphics[scale=0.7]{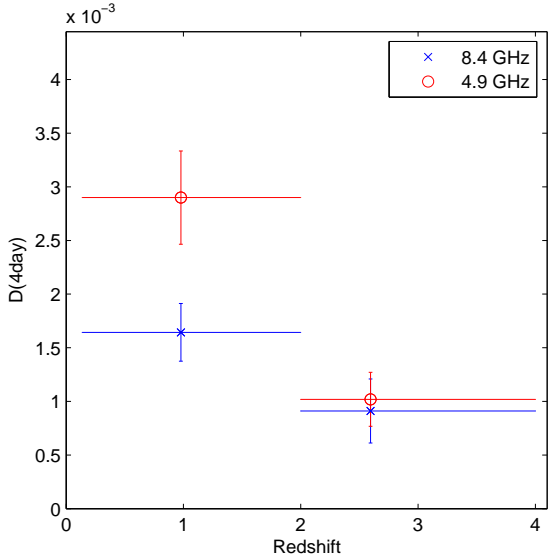}
     \end{center}
     \caption{{Mean 4-day Structure Functions at 8.4 GHz and 4.9 GHz in two redshift bins} \label{redshift2bins}}
     \end{figure}

Figure~\ref{redshift} shows scatter and binned plots of $D(\rm 4day)$ against source redshift at 8.4 GHz and 4.9 GHz. The redshift dependence of AGN variability is evident at both frequencies, confirming the result of the MASIV survey. We obtained Kendall's tau rank correlation coefficients of -0.34 at 8.4 GHz and -0.33 at 4.9 GHz with p-values of $1.2 \times 10^{-8}$ and $2.2 \times 10^{-8}$. Although not obvious from the scatter plots and from the rank correlation coefficients, there appears to be a frequency dependence in the scaling of the mean $D(\rm 4day)$ with redshift. This can be seen in the binned plots in Figure~\ref{redshift}, and in Figure~\ref{redshift2bins} where the sources are grouped into just two redshift bins. The mean 4.9 GHz $D(\rm 4day)$ at $z > 2$ is about a factor of 3 lower than its $z < 2$ counterpart. On the other hand, the 8.4 GHz $D(\rm 4day)$ drops only by about a factor of 1.8 from low to high redshift.

In the limit of weak interstellar scintillation resulting from a thin scattering screen with Kolmogorov spectrum turbulence, the modulation index of a point source, $m_{p}$, is given by \citep{walker98}:
\begin{equation}\label{mp}
m_{p} = \left( \dfrac{\lambda}{\lambda_{t}}\right)^{17/12},
\end{equation}
where $\lambda$ is the observing wavelength and $\lambda_{t}$ is the transition wavelength between weak and strong ISS.
The observed modulation index of an extended source, $m_{obs}$, is suppressed relative to that of a point source and is given by \citep{walker98}:
\begin{equation}\label{mobs}
m_{obs} = m_{p}\left( \dfrac{\theta_{F}}{\theta_{obs}}\right)^{7/6},
\end{equation}
where $\theta_{F} = \sqrt{\lambda/2 \pi L}$ is the angular size of the first Fresnel zone at the scattering screen of the ISM, and $L$ is the distance between the Earth and the scattering screen. $\theta_{obs}$ is the angular size of the source, which represents its intrinsic size and an additional increase in diameter due to scatter broadening in the IGM so that $\theta_{obs}^{2} = \theta_{int}^{2} + \theta_{igm}^{2}$. The intrisic source size can be modeled as:
\begin{equation}\label{brightnesstemp}
\theta_{int}^{2}=\sqrt{\dfrac{\lambda^{2}(1+z) S}{2 \pi k \delta T_{b}}},
\end{equation}
where $S$ is the observed mean flux density of the source, $\delta$ is the Doppler boosting factor, $T_{b}$ is the intrinsic brightness temperature of the source, and the $(1+z)$ term accounts for the effect of cosmological expansion when converting the source brightness temperature in the emission frame to the observer's frame. We also know that $\theta_{igm} \propto \lambda^{2}$. At low redshifts, we expect $\theta_{int}$ to dominate, so that $\theta_{obs} \propto \lambda$. Substituting this into Equation~\ref{mobs} and making use of Equation~\ref{mp}, we find that the $m_{obs}$ at 4.9 GHz ($\lambda$ = 6 cm) should be about a factor of 1.5 larger than $m_{obs}$ at 8.4 GHz ($\lambda$ = 3.6 cm), assuming that the Doppler boosting factor and the source brightness temperature are frequency independent. Since we are interested in the SFs (which when saturated is $\approx 2m_{obs}^{2}$), this factor should be $\approx 2.25$. If there is no scatter broadening in the IGM, this factor should remain unchanged at high redshift, even if the source Doppler factors evolve with redshift. However, if scatter broadening in the IGM dominates at high redshift, $\theta_{obs} \propto \lambda^{2}$, and the ratio of $D(\rm 4day)$ at 4.9 GHz to that at 8.4 GHz is estimated from the model to be $\approx 0.7$ (0.8 for the ratio of $m_{obs}$). As it is unlikely for $\theta_{obs}$ to be entirely dominated by IGM scatter broadening at high redshift, we expect the mean 8.4 GHz $D(\rm 4day)$ to be at least comparable, if not larger, than that at 4.9 GHz. 

Figure~\ref{redshift2bins} clearly shows a reduction in the ratio of the 4.9 GHz $D(\rm 4day)$ to the 8.4 GHz $D(\rm 4day)$ from $\approx$ 1.8 at $z < 2$ to $\approx$ 1.1 at $z > 2$. As the mean values of $D(\rm 4day)$ at $z > 2$ for both frequencies are an order of magnitude larger than the lower limit of measureable variability, we know that this effect is not a result of the mean SFs at both frequencies hitting the noise floor. Also, the model calculations show that this frequency scaling of the redshift dependence is expected to be weak, with a factor of 2 to 3 reduction in the SF ratios from low to high redshift. This may explain why this frequency scaling is not discernible from the log scale scatter plots. The two-sample Kolmogorov-Smirnov (K-S) test rejects the null hypothesis that the distributions of the 4.9 GHz $D(\rm 4day)$ and the 8.4 GHz $D(\rm 4day)$ at $z < 2$ are drawn from the same parent population at a significance level of 0.05 (with a p-value of 0.01). However, at $z > 2$, the K-S Test no longer gives a statistically significant rejection of the same null hypothesis, with a p-value of 0.26. While this in no way proves that the distributions of the 4.9 GHz $D(\rm 4day)$ and the 8.4 GHz $D(\rm 4day)$ are similar at high redshift, it is still an interesting result.

Although the results appear tantalizing, we note that a combination of various selection effects, including source spectral indices (as demonstrated in Section~\ref{spectralindexsec}) and luminosities \citep{bignalletal10}, can lead to spurious interpretations. Furthermore, the above calculations are based on the assumption of weak ISS at both frequencies, whereas the transition frequency between weak and strong scattering is close to 5 GHz at mid-Galactic latitudes where the sources in our sample lie. There is also a possibility that the transition frequency may be higher than 5 GHz for some lines of sight where the sources are seen through thicker regions of the Galaxy. It is therefore crucial to understand why the 8.4 GHz $D(\rm 4day)$ is comparable to or larger than the 4.9 GHz $D(\rm 4 day)$ in $\approx 35\%$ of the sources as seen in Figure~\ref{SFvsSF}. As discussed in Section~\ref{individual} for the source J1919+3159, this effect can be a result of increased scatter broadening at 4.9 GHz relative to 8.4 GHz, leading to an increase in apparent source size at 4.9 GHz, or due to the presence of strong refractive scintillation in these sources. Any bias towards strong scattering (or large transition frequencies) in the high redshift sample will affect the interpretation of Figure~\ref{redshift2bins}. Therefore, while much insight can be gained from using the weak scattering approximation, the observations will need to be compared with models based on numerical computations at intermediate scintillation regimes where no analytical formulae exist. We thus defer a full discussion of all these complicating effects and further interpretation to a follow-up paper.

   \section{Conclusion}\label{conclusion}
Multi-frequency observations of 140 flat spectrum AGN were carried out using the VLA for a total duration of 11 days. The sensitivity of the VLA and careful calibration enabled noise and systematic errors to be reduced down to a level of $\lesssim 1\%$. These errors were then quantified as a quadratic sum of the noise ($s$) and calibration errors ($p$), and subsequently subtracted from the SF values obtained by fitting a simple model to the SFs calculated from the source lightcurves. Statistically significant correlations with H$\alpha$ intensities were obtained for the SFs at all time-lags (using both model and single sample estimates calculated from the lightcurves) and at both frequencies, confirming the MASIV results linking the variability of the sources to ISS. Cross-covariance functions between source lightcurves at 4.9 GHz and 8.4 GHz reveal that the patterns of scintillating sources are correlated. A time-lag shift in the peak of the cross-covariance function points to the possibility of a core shift in such sources at different frequencies. Although there were no clear trends with regards to mean spectral indices above -0.4, a clear drop in ISS amplitudes was observed for sources with spectral indices below -0.4, confirming reduced scintillation in steeper spectrum sources. As in the MASIV survey, a drop in ISS at $z \gtrsim 2$ was observed at 4.9 GHz. Of even greater significance is the detected reduction in the redshift dependence of ISS at 8.4 GHz, suggestive of scatter broadening in the IGM if weak ISS is assumed. A follow-up paper will delve further into its interpretation, pending a full investigation into source selection effects and comparisons with more accurate models to understand the complex underlying physics.

The results of MASIV and this study continue to demonstrate the potential of using ISS as an astrophysical and cosmological probe. As shown, ISS can be used to estimate the core-shifts of radio sources at a higher resolution than that of VLBI. Such observations will be important for astrometric applications in the selection of sources for an International Celestial Reference Frame. Combining multi-frequency scintillation observations with VLBI imaging (to obtain the true angular separation between the cores) raises the prospect of putting constraints on actual scattering screen distances and velocities, providing further insight into the physics of the ISM.

While this study included only 140 sources, future large scale surveys of IDV are already being planned, such as the VAST Survey \citep{murphyandchatterjee09} - one of the key survey science projects of the ASKAP. These future experiments using the ASKAP will operate at a much higher survey speed. Thus the various techniques applied in this study, in our effort to obtain the best possible characterization of the variability of the sources, provide valuable insight for these future surveys, which require the development of efficient pipelined algorithms for the calibration, detection and analysis of the large quantities of data.
   
   \section{Acknowledgments}
J. Y. Koay is supported by the Curtin Strategic International Research Scholarship (CSIRS) provided by Curtin University, and would like to thank Prof. Ron Ekers of CSIRO and Greg Taylor of the University of New Mexico for helpful discussions. B. J. Rickett thanks the National Science Foundation (NSF) for partial support under grant AST-0507713. R. Ojha is supported by an appointment to the NASA Postdoctoral Program at the Goddard Space Flight Center, administered by Oak Ridge Associated Universities through a contract with NASA. We all thank the operators and scientific staff at the Very Large Array; in particular we thank Vivek Dhawan for his extensive advice and help during our long sequence of observations. The VLA is part of the National Radio Astronomy Observatory (NRAO), which is a facility of the NSF operated under cooperative agreement by Associated Universities, Inc. This study made use of data from the Wisconsin H-Alpha Mapper (WHAM) survey, which is funded by the NSF. Data was also obtained from the NASA/IPAC Extragalactic Database (NED), which is operated by the Jet Propulsion Laboratory, California Institute of Technology, under contract with NASA.

   \appendix
   \section{Appendix}
   \input{AL730list4.tex}

\end{document}

%% file: freqcorrtab.tex
\begin{table}[!htp]
\caption{{List of sources where the cross-covariance function of the 4.9 and 8.4 GHz lightcurves peak at non-zero timelags.} \label{freqcorrtable}}
\scriptsize
\begin{tabular}{c r @{ $\pm$ } l}
\hline
Source Name & \multicolumn{2}{c}{Timelag (Days)} \\ 
\hline
\hline

J0017+5312 & -0.43 & 0.05 \\
J0154+4743 &  2.02 & 0.06 \\
J0308+1208 &  0.79 & 0.04 \\
J0342+3859 &  0.95 & 0.04 \\
J0409+1217 & -1.08 & 0.04 \\
J0449+1121 &  2.89 & 0.03 \\
J0510+1800 &  0.24 & 0.04 \\
J0659+0813 &  0.24 & 0.03 \\
J0726+6125 & -0.32 & 0.04 \\
J0741+2557 & -0.16 & 0.04 \\
J0750+1231 &  0.80 & 0.03 \\
J0757+0956 &  1.05 & 0.03 \\
J0825+0309 &  0.21 & 0.03 \\
J1410+6141 & -0.40 & 0.04 \\
J1417+3818 &  1.59 & 0.04 \\
J1535+6953 &  0.42 & 0.05 \\
J1701+0338 &  0.21 & 0.03 \\
J1734+3857 &  0.19 & 0.03 \\
J1800+3848 &  2.04 & 0.04 \\
J1905+1943 & -0.26 & 0.03 \\
J1919+3159 & -1.32 & 0.02 \\
J2012+6319 & -0.37 & 0.05 \\
J2113+1121 &  1.20 & 0.04 \\
J2237+4216 &  0.31 & 0.04 \\
J2253+3236 &  1.91 & 0.04 \\
\hline
\end{tabular}

\end{table}

%% file: AL730list4.tex
\begin{deluxetable}{c c c c c c c c c c c}

\tabletypesize{\scriptsize}
\tablecaption{List of Sources and Relevant Information\label{AL730list}}
\tablehead{
\colhead{Source} & \colhead{H$\alpha$} & \colhead{$S_{8.4}$} & \colhead{$S_{4.9}$} & \colhead{$\alpha$} & \colhead{$\tau_{char,8.4}$} & \colhead{$\tau_{char,4.9}$} & \colhead{$m_{8.4}$} & \colhead{$m_{4.9}$} & \colhead{$D_{8.4}(4day)$} & \colhead{$D_{4.9}(4day)$}\\

\colhead{(1)} & \colhead{(2)} & \colhead{(3)} & \colhead{(4)} & \colhead{(5)} & \colhead{(6)} & \colhead{(7)} & \colhead{(8)} & \colhead{(9)} & \colhead{(10)} & \colhead{(11)}
}

\startdata
J0009+1513 &     0.7 &    0.12 &    0.15 &   -0.45 &  -  &  -  &   1.5 &   1.2 & 2.40E-004 $\pm$ 1.8E-004 & 1.89E-004 $\pm$ 1.5E-004 \\ 
J0017+5312 &    13.3 &    0.64 &    0.59 &    0.13 &  -  & $>$ 11 &   1.1 &   1.6 & 2.42E-004 $\pm$ 3.1E-005 & 4.68E-004 $\pm$ 1.6E-004 \\ 
J0017+8135 &     2.2 &    1.26 &    1.36 &   -0.13 &  -  &  -  &   0.6 &   0.6 & 5.10E-006 $\pm$ 2.8E-006 & 1.89E-005 $\pm$ 1.5E-005 \\ 
J0056+1625 &     0.8 &    0.23 &    0.19 &    0.33 & 0.4$\pm$0.3 & 0.3$\pm$0.1 &   3.4 &   3.3 & 2.21E-003 $\pm$ 5.1E-004 & 2.40E-003 $\pm$ 8.1E-004 \\ 
J0108+0135 &     0.7 &    2.07 &    1.53 &    0.56 &  -  & $>$ 11 &   0.9 &   1.9 & 1.13E-004 $\pm$ 5.1E-005 & 6.49E-004 $\pm$ 1.9E-004 \\ 
J0122+0310 &     0.5 &    0.11 &    0.11 &   -0.04 & 0.5$\pm$0.5 & 0.3$\pm$0.4 &   1.6 &   1.7 & 3.17E-004 $\pm$ 1.1E-004 & 3.19E-004 $\pm$ 1.5E-004 \\ 
J0122+2502 &     0.9 &    0.66 &    0.75 &   -0.21 &  -  &  -  &   0.5 &   0.9 & 3.03E-007 $\pm$ 9.3E-005 & 6.65E-005 $\pm$ 4.6E-004 \\ 
J0126+2559 &     1.0 &    0.66 &    0.81 &   -0.39 &  -  &  -  &   0.5 &   0.9 & 5.91E-006 $\pm$ 5.7E-005 & 6.85E-005 $\pm$ 4.6E-005 \\ 
J0135+2158 &     0.9 &    0.14 &    0.18 &   -0.37 &  -  & 0.5$\pm$0.2 &   1.4 &   1.9 & 2.75E-004 $\pm$ 1.1E-004 & 1.02E-003 $\pm$ 3.7E-004 \\ 
J0154+4743 &     8.6 &    0.60 &    0.50 &    0.35 & 0.1$\pm$0.0 & 0.5$\pm$0.4 &   2.2 &   2.0 & 9.43E-004 $\pm$ 1.5E-004 & 7.53E-004 $\pm$ 1.6E-004 \\ 
J0217+7349 &     2.2 &    4.21 &    4.31 &   -0.05 &  -  &  -  &   0.3 &   0.5 & 4.96E-006 $\pm$ 1.1E-005 & 4.50E-007 $\pm$ 1.4E-005 \\ 
J0237+2046 &     1.3 &    0.13 &    0.13 &   -0.02 & 0.3$\pm$0.3 & 0.2$\pm$0.1 &   1.8 &   2.1 & 5.31E-004 $\pm$ 6.4E-005 & 8.65E-004 $\pm$ 2.7E-004 \\ 
J0238+1636 &     1.1 &    3.60 &    3.66 &   -0.03 & $>$ 11 & $>$ 11 &   2.5 &   2.4 & 1.44E-003 $\pm$ 3.1E-004 & 1.16E-003 $\pm$ 3.6E-004 \\ 
J0242+1101 &     1.0 &    0.82 &    0.92 &   -0.22 &  -  &  -  &   1.2 &   1.3 & 2.09E-004 $\pm$ 5.0E-005 & 2.62E-004 $\pm$ 1.2E-004 \\ 
J0259+1925 &     1.3 &    0.18 &    0.14 &    0.49 & 0.7$\pm$0.5 & $>$ 11 &   3.8 &   5.5 & 3.28E-003 $\pm$ 7.0E-004 & 4.85E-003 $\pm$ 1.5E-003 \\ 
J0308+1208 &     1.9 &    0.06 &    0.07 &   -0.17 & 0.2$\pm$0.0 & 1.4$\pm$0.7 &   2.8 &   2.5 & 1.28E-003 $\pm$ 3.1E-004 & 1.29E-003 $\pm$ 4.7E-004 \\ 
J0313+0228 &     6.9 &    0.10 &    0.12 &   -0.23 & 9.6$\pm$7.7 & $>$ 11 &   3.9 &   6.8 & 3.52E-003 $\pm$ 4.3E-004 & 9.36E-003 $\pm$ 1.8E-003 \\ 
J0321+1221 &     2.1 &    1.52 &    1.68 &   -0.19 &  -  &  -  &   1.1 &   0.9 & 1.72E-004 $\pm$ 7.3E-005 & 4.03E-005 $\pm$ 6.3E-006 \\ 
J0323+0446 &     4.6 &    0.14 &    0.12 &    0.16 & 0.8$\pm$0.3 & 0.6$\pm$0.2 &   3.3 &   5.6 & 2.13E-003 $\pm$ 4.7E-004 & 7.27E-003 $\pm$ 2.3E-003 \\ 
J0342+3859 &     8.1 &    0.10 &    0.09 &    0.27 & $>$ 11 & $>$ 11 &   7.6 &   5.7 & 1.16E-002 $\pm$ 1.5E-003 & 5.77E-003 $\pm$ 1.3E-003 \\ 
J0343+3622 &     7.8 &    0.28 &    0.32 &   -0.22 & 1.0$\pm$0.9 & 2.9$\pm$2.8 &   2.4 &   2.8 & 1.00E-003 $\pm$ 1.5E-004 & 1.34E-003 $\pm$ 2.2E-004 \\ 
J0358+3850 &    12.3 &    0.18 &    0.18 &   -0.00 & 1.2$\pm$1.2 & 9.9$\pm$1.5 &   4.2 &   4.1 & 4.02E-003 $\pm$ 6.1E-004 & 4.28E-003 $\pm$ 1.0E-003 \\ 
J0403+2600 &     4.7 &    2.17 &    2.13 &    0.03 &  -  &  -  &   0.9 &   0.8 & 1.07E-004 $\pm$ 2.5E-005 & 2.89E-005 $\pm$ 1.8E-005 \\ 
J0406+2511 &     4.5 &    0.11 &    0.12 &   -0.15 & 1.9$\pm$1.4 & $>$ 11 &   3.6 &   8.5 & 1.93E-003 $\pm$ 3.3E-004 & 1.49E-002 $\pm$ 2.7E-003 \\ 
J0409+1217 &     7.6 &    0.29 &    0.32 &   -0.19 & 9.6$\pm$4.4 & 4.5$\pm$3.3 &   5.0 &   4.0 & 4.60E-003 $\pm$ 7.8E-004 & 3.52E-003 $\pm$ 6.3E-004 \\ 
J0422+0219 &     4.6 &    1.07 &    1.25 &   -0.29 & 0.2$\pm$0.1 & 0.7$\pm$0.9 &   1.4 &   1.5 & 3.37E-004 $\pm$ 8.4E-005 & 4.12E-004 $\pm$ 2.5E-004 \\ 
J0449+1121 &     9.8 &    0.77 &    0.86 &   -0.19 & 0.6$\pm$0.5 & $>$ 11 &   1.6 &   2.3 & 6.00E-004 $\pm$ 1.1E-004 & 6.22E-004 $\pm$ 2.3E-004 \\ 
J0459+0229 &     7.4 &    0.72 &    1.06 &   -0.72 &  -  &  -  &   1.2 &   1.0 & 2.45E-004 $\pm$ 4.6E-005 & 8.63E-005 $\pm$ 5.4E-005 \\ 
J0510+1800 &    22.1 &    0.96 &    0.73 &    0.52 & 0.5$\pm$0.1 & 0.5$\pm$0.7 &   6.3 &   7.7 & 7.84E-003 $\pm$ 2.0E-003 & 7.89E-003 $\pm$ 4.6E-003 \\ 
J0530+1331 &    70.5 &    3.19 &    3.30 &   -0.06 & 2.7$\pm$2.4 & $>$ 11 &   6.9 &   5.6 & 1.12E-002 $\pm$ 1.4E-003 & 7.10E-003 $\pm$ 1.7E-003 \\ 
J0534+1047 &   147.0 &    0.16 &    0.17 &   -0.13 &  -  &  -  &   0.4 &   0.7 & 3.66E-006 $\pm$ 1.7E-005 & 5.70E-006 $\pm$ 1.2E-005 \\ 
J0539+1433 &    30.3 &    0.43 &    0.37 &    0.29 & 0.4$\pm$0.3 & 0.2$\pm$0.4 &   4.0 &   3.1 & 3.23E-003 $\pm$ 6.2E-004 & 1.52E-003 $\pm$ 7.1E-004 \\ 
J0614+6046 &     1.7 &    0.49 &    0.70 &   -0.66 &  -  &  -  &   1.0 &   0.9 & 1.51E-004 $\pm$ 3.0E-005 & 4.87E-005 $\pm$ 2.3E-005 \\ 
J0624+3856 &     1.7 &    0.70 &    0.87 &   -0.40 & 0.8$\pm$0.5 & $>$ 11 &   1.3 &   2.2 & 3.34E-004 $\pm$ 6.3E-005 & 1.12E-003 $\pm$ 2.4E-004 \\ 
J0646+4451 &     0.9 &    3.67 &    3.06 &    0.33 &  -  &  -  &   0.3 &   0.6 & 5.68E-006 $\pm$ 1.3E-005 & 8.58E-007 $\pm$ 1.3E-005 \\ 
J0659+0813 &     7.6 &    0.67 &    0.73 &   -0.16 & 0.6$\pm$0.4 & 0.9$\pm$0.1 &   1.4 &   1.6 & 4.00E-004 $\pm$ 7.3E-005 & 3.69E-004 $\pm$ 1.2E-004 \\ 
J0726+6125 &     1.0 &    0.37 &    0.24 &    0.80 & $>$ 11 & 1.4$\pm$1.3 &   2.3 &   3.0 & 1.10E-003 $\pm$ 8.2E-005 & 1.85E-003 $\pm$ 2.6E-004 \\ 
J0739+0137 &     1.8 &    1.94 &    1.74 &    0.20 & 0.7$\pm$0.4 & 7.9$\pm$9.3 &   1.5 &   3.7 & 4.53E-004 $\pm$ 9.9E-005 & 3.09E-003 $\pm$ 1.3E-003 \\ 
J0739+7527 &     1.4 &    0.14 &    0.22 &   -0.80 &  -  &  -  &   0.9 &   1.2 & 6.90E-005 $\pm$ 1.2E-005 & 2.70E-004 $\pm$ 4.7E-005 \\ 
J0741+2557 &     2.8 &    0.08 &    0.05 &    0.83 & 0.4$\pm$0.1 & 0.1$\pm$0.1 &   1.6 &   3.4 & 4.27E-004 $\pm$ 1.1E-004 & 1.85E-003 $\pm$ 4.9E-004 \\ 
J0745+1011 &     1.4 &    2.06 &    2.95 &   -0.67 &  -  &  -  &   0.7 &   0.6 & 2.96E-005 $\pm$ 1.5E-005 & 1.10E-006 $\pm$ 2.4E-005 \\ 
J0750+1231 &     1.8 &    4.16 &    3.76 &    0.19 &  -  &  -  &   1.2 &   1.6 & 2.69E-004 $\pm$ 4.2E-005 & 2.91E-004 $\pm$ 1.7E-004 \\ 
J0757+0956 &     1.5 &    1.10 &    1.02 &    0.14 & 0.2$\pm$0.0 & 0.2$\pm$0.1 &   2.7 &   3.6 & 1.44E-003 $\pm$ 2.8E-004 & 2.15E-003 $\pm$ 4.8E-004 \\ 
J0800+4854 &     0.5 &    0.08 &    0.10 &   -0.39 & 0.0$\pm$0.0 & 0.1$\pm$0.1 &   5.1 &   5.0 & 4.67E-003 $\pm$ 5.0E-004 & 4.74E-003 $\pm$ 6.2E-004 \\ 
J0805+6144 &     0.8 &    0.72 &    0.97 &   -0.56 &  -  &  -  &   0.4 &   0.6 & 2.04E-006 $\pm$ 1.0E-005 & 6.15E-006 $\pm$ 2.9E-006 \\ 
J0810+1010 &     1.3 &    0.09 &    0.11 &   -0.31 &  -  &  -  &   1.0 &   1.5 & 9.16E-005 $\pm$ 3.9E-005 & 1.43E-004 $\pm$ 1.4E-004 \\ 
J0811+0146 &     1.7 &    0.98 &    0.65 &    0.76 & 1.4$\pm$0.5 & 2.4$\pm$2.2 &   2.2 &   4.3 & 8.09E-004 $\pm$ 1.9E-004 & 5.27E-003 $\pm$ 9.2E-004 \\ 
J0818+4222 &     1.2 &    1.41 &    1.35 &    0.08 & 0.1$\pm$0.1 & 0.1$\pm$0.2 &   2.2 &   3.8 & 8.16E-004 $\pm$ 8.2E-005 & 2.55E-003 $\pm$ 5.7E-004 \\ 
J0821+3107 &     2.3 &    0.06 &    0.08 &   -0.66 &  -  & 0.3$\pm$0.1 &   1.3 &   2.3 & 1.49E-004 $\pm$ 3.4E-005 & 8.51E-004 $\pm$ 2.1E-004 \\ 
J0825+0309 &     1.0 &    1.53 &    1.38 &    0.19 &  -  & 0.4$\pm$0.3 &   1.1 &   2.3 & 1.59E-004 $\pm$ 5.0E-005 & 8.73E-004 $\pm$ 2.2E-004 \\ 
J0850+5159 &     0.7 &    0.08 &    0.10 &   -0.34 &  -  &  -  &   0.9 &   1.2 & 3.73E-005 $\pm$ 1.1E-005 & 1.45E-004 $\pm$ 3.4E-005 \\ 
J0854+8034 &     1.6 &    0.22 &    0.25 &   -0.19 &  -  &  -  &   0.8 &   1.0 & 6.72E-005 $\pm$ 1.4E-005 & 1.25E-004 $\pm$ 2.3E-005 \\ 
J0856+7146 &     0.9 &    0.11 &    0.07 &    0.83 & $>$ 11 & 0.7$\pm$0.4 &   4.0 &   5.4 & 2.58E-003 $\pm$ 3.5E-004 & 4.71E-003 $\pm$ 9.6E-004 \\ 
J0914+0245 &     2.1 &    0.89 &    1.08 &   -0.36 &  -  &  -  &   0.8 &   1.1 & 7.53E-005 $\pm$ 3.3E-005 & 2.12E-004 $\pm$ 7.2E-005 \\ 
J0916+0242 &     1.9 &    0.11 &    0.09 &    0.39 & 0.3$\pm$0.2 & 0.3$\pm$0.1 &   4.4 &   7.7 & 3.30E-003 $\pm$ 6.1E-004 & 9.91E-003 $\pm$ 4.3E-003 \\ 
J0920+4441 &     0.4 &    1.34 &    1.09 &    0.38 &  -  &  -  &   0.6 &   0.8 & 1.82E-005 $\pm$ 1.0E-005 & 8.13E-005 $\pm$ 3.4E-005 \\ 
J0929+5013 &     0.6 &    0.39 &    0.40 &   -0.06 & 0.0$\pm$0.1 & 0.3$\pm$0.2 &   2.6 &   4.1 & 1.22E-003 $\pm$ 1.0E-004 & 3.03E-003 $\pm$ 5.1E-004 \\ 
J0953+1720 &     0.9 &    0.07 &    0.10 &   -0.67 & 0.9$\pm$0.2 & 0.1$\pm$0.1 &   1.6 &   2.0 & 3.93E-004 $\pm$ 8.6E-005 & 6.12E-004 $\pm$ 1.5E-004 \\ 
J0958+4725 &     0.7 &    1.26 &    1.52 &   -0.36 &  -  &  -  &   0.8 &   0.8 & 7.57E-005 $\pm$ 1.5E-005 & 5.77E-005 $\pm$ 2.2E-005 \\ 
J0958+6533 &     1.1 &    0.98 &    1.07 &   -0.16 & $>$ 11 & 6.6$\pm$2.5 &   1.7 &   1.8 & 4.84E-004 $\pm$ 4.1E-005 & 5.30E-004 $\pm$ 9.5E-005 \\ 
J1007+1356 &     1.1 &    0.66 &    0.71 &   -0.12 &  -  &  -  &   0.7 &   0.9 & 2.33E-005 $\pm$ 1.1E-005 & 6.38E-005 $\pm$ 4.4E-005 \\ 
J1016+2037 &     0.8 &    0.46 &    0.61 &   -0.53 &  -  &  -  &   0.5 &   0.7 & 1.32E-007 $\pm$ 1.4E-005 & 5.02E-005 $\pm$ 4.4E-005 \\ 
J1049+1429 &     0.9 &    0.15 &    0.13 &    0.22 & 0.0$\pm$0.1 & 0.2$\pm$0.1 &   2.8 &   3.0 & 1.41E-003 $\pm$ 1.4E-004 & 1.58E-003 $\pm$ 4.2E-004 \\ 
J1056+7011 &    -0.2 &    0.35 &    0.28 &    0.40 & 10.2$\pm$8.4 & 0.6$\pm$0.5 &   1.8 &   2.3 & 6.39E-004 $\pm$ 4.5E-005 & 8.90E-004 $\pm$ 1.1E-004 \\ 
J1125+2610 &     0.5 &    1.02 &    1.17 &   -0.26 &  -  &  -  &   0.7 &   0.9 & 4.19E-005 $\pm$ 1.1E-005 & 9.09E-005 $\pm$ 3.0E-005 \\ 
J1159+2914 &     0.5 &    3.23 &    2.60 &    0.40 & 0.1$\pm$0.1 & 0.2$\pm$0.0 &   1.9 &   3.6 & 6.48E-004 $\pm$ 7.5E-005 & 2.63E-003 $\pm$ 6.2E-004 \\ 
J1247+7046 &     0.4 &    0.11 &    0.09 &    0.36 & 0.3$\pm$0.2 & 0.3$\pm$0.2 &   2.0 &   2.9 & 7.39E-004 $\pm$ 6.9E-005 & 1.72E-003 $\pm$ 2.5E-004 \\ 
J1316+6927 &     0.4 &    0.11 &    0.12 &   -0.20 &  -  &  -  &   1.0 &   1.5 & 1.11E-004 $\pm$ 1.8E-005 & 2.83E-004 $\pm$ 7.6E-005 \\ 
J1328+6221 &     0.5 &    0.08 &    0.10 &   -0.26 & 0.0$\pm$0.0 & 0.0$\pm$0.1 &   3.0 &   5.9 & 1.63E-003 $\pm$ 1.4E-004 & 6.29E-003 $\pm$ 7.6E-004 \\ 
J1330+4954 &     0.5 &    0.09 &    0.11 &   -0.35 &  -  &  -  &   0.8 &   1.2 & 4.18E-005 $\pm$ 1.2E-005 & 1.53E-004 $\pm$ 4.3E-005 \\ 
J1354+6645 &     0.6 &    0.08 &    0.08 &    0.01 & 0.3$\pm$0.3 & 0.2$\pm$0.1 &   3.2 &   2.3 & 1.91E-003 $\pm$ 2.3E-004 & 8.56E-004 $\pm$ 1.3E-004 \\ 
J1410+6141 &     0.4 &    0.18 &    0.17 &    0.16 &  -  & $>$ 11 &   1.1 &   2.0 & 1.91E-004 $\pm$ 2.2E-005 & 6.53E-004 $\pm$ 8.7E-005 \\ 
J1417+3818 &     0.4 &    0.10 &    0.12 &   -0.32 & $>$ 11 & 0.6$\pm$0.6 &   2.1 &   2.7 & 8.33E-004 $\pm$ 7.0E-005 & 1.32E-003 $\pm$ 2.1E-004 \\ 
J1436+6336 &     0.5 &    1.65 &    1.42 &    0.28 &  -  &  -  &   0.6 &   0.9 & 2.26E-007 $\pm$ 9.0E-006 & 5.18E-005 $\pm$ 3.9E-005 \\ 
J1437+5112 &     0.7 &    0.08 &    0.11 &   -0.53 & 0.1$\pm$0.0 & 0.6$\pm$0.3 &   1.8 &   2.1 & 4.94E-004 $\pm$ 6.0E-005 & 6.95E-004 $\pm$ 1.4E-004 \\ 
J1442+0625 &     0.6 &    0.08 &    0.08 &   -0.04 & 1.2$\pm$0.7 & 0.1$\pm$0.0 &   3.3 &   3.8 & 1.51E-003 $\pm$ 1.6E-004 & 2.41E-003 $\pm$ 5.8E-004 \\ 
J1535+4836 &     0.2 &    0.15 &    0.14 &    0.10 &  -  &  -  &   0.8 &   0.9 & 4.54E-005 $\pm$ 1.4E-005 & 2.61E-005 $\pm$ 1.9E-005 \\ 
J1549+5038 &     0.5 &    0.93 &    0.91 &    0.04 & 0.0$\pm$0.1 & 0.0$\pm$0.1 &   2.1 &   3.0 & 7.71E-004 $\pm$ 8.8E-005 & 1.62E-003 $\pm$ 3.2E-004 \\ 
J1559+0805 &     0.9 &    0.15 &    0.12 &    0.39 &  -  &  -  &   0.9 &   1.1 & 7.66E-005 $\pm$ 1.8E-005 & 1.24E-004 $\pm$ 3.9E-005 \\ 
J1610+7809 &     0.8 &    0.15 &    0.18 &   -0.34 &  -  & 0.4$\pm$0.4 &   1.3 &   2.6 & 2.74E-004 $\pm$ 2.8E-005 & 1.26E-003 $\pm$ 1.8E-004 \\ 
J1616+0459 &     1.1 &    0.81 &    0.93 &   -0.25 &  -  &  -  &   0.9 &   0.7 & 1.26E-004 $\pm$ 3.0E-005 & 2.00E-006 $\pm$ 1.9E-005 \\ 
J1619+2247 &     0.6 &    0.68 &    0.70 &   -0.07 &  -  &  -  &   1.0 &   1.3 & 1.25E-004 $\pm$ 2.3E-005 & 2.12E-004 $\pm$ 4.0E-005 \\ 
J1625+4134 &     0.4 &    0.84 &    1.04 &   -0.40 &  -  &  -  &   0.7 &   0.9 & 3.93E-005 $\pm$ 8.3E-006 & 5.23E-005 $\pm$ 3.0E-005 \\ 
J1639+4128 &     0.5 &    0.15 &    0.13 &    0.25 & 0.6$\pm$0.5 & $>$ 11 &   1.8 &   3.5 & 4.85E-004 $\pm$ 6.7E-005 & 1.88E-003 $\pm$ 3.0E-004 \\ 
J1659+1714 &     1.1 &    0.11 &    0.13 &   -0.26 &  -  & $>$ 11 &   1.2 &   1.9 & 2.05E-004 $\pm$ 4.0E-005 & 5.46E-004 $\pm$ 1.5E-004 \\ 
J1701+0338 &     2.0 &    0.09 &    0.11 &   -0.38 & 4.4$\pm$4.4 & $>$ 11 &   1.8 &   3.8 & 6.95E-004 $\pm$ 8.6E-005 & 2.44E-003 $\pm$ 8.2E-004 \\ 
J1716+6836 &     1.3 &    0.61 &    0.54 &    0.22 &  -  &  -  &   1.0 &   1.0 & 1.40E-004 $\pm$ 2.6E-005 & 9.73E-005 $\pm$ 4.2E-005 \\ 
J1719+0817 &     2.0 &    0.49 &    0.59 &   -0.34 &  -  &  -  &   0.8 &   1.3 & 5.98E-005 $\pm$ 1.3E-005 & 2.56E-004 $\pm$ 8.7E-005 \\ 
J1719+1745 &     1.1 &    0.60 &    0.63 &   -0.11 & 2.1$\pm$1.9 & $>$ 11 &   2.3 &   2.5 & 1.14E-003 $\pm$ 1.3E-004 & 1.15E-003 $\pm$ 2.2E-004 \\ 
J1728+0427 &     2.4 &    0.46 &    0.47 &   -0.05 & 1.4$\pm$0.9 & 8.7$\pm$0.3 &   2.3 &   1.6 & 1.22E-003 $\pm$ 2.4E-004 & 5.04E-004 $\pm$ 1.2E-004 \\ 
J1733+1635 &     1.4 &    0.07 &    0.11 &   -0.77 &  -  &  -  &   1.2 &   1.3 & 1.35E-004 $\pm$ 3.7E-005 & 1.55E-004 $\pm$ 5.1E-005 \\ 
J1734+3857 &     1.4 &    0.82 &    0.82 &    0.00 & $>$ 11 & 0.4$\pm$0.2 &   2.2 &   2.0 & 8.92E-004 $\pm$ 7.3E-005 & 4.99E-004 $\pm$ 2.1E-004 \\ 
J1740+5211 &     0.9 &    1.12 &    0.99 &    0.24 & 0.5$\pm$0.4 & 0.5$\pm$0.3 &   1.4 &   1.5 & 3.66E-004 $\pm$ 3.7E-005 & 3.48E-004 $\pm$ 7.3E-005 \\ 
J1742+5945 &     1.3 &    0.19 &    0.19 &    0.00 &  -  & 1.5$\pm$1.3 &   1.2 &   3.7 & 2.02E-004 $\pm$ 2.6E-005 & 2.78E-003 $\pm$ 4.1E-004 \\ 
J1745+4059 &     2.1 &    0.11 &    0.10 &    0.14 & 0.8$\pm$0.7 & 4.2$\pm$3.2 &   2.6 &   6.4 & 1.36E-003 $\pm$ 1.5E-004 & 9.01E-003 $\pm$ 1.1E-003 \\ 
J1751+0939 &     3.1 &    5.22 &    3.88 &    0.55 & $>$ 11 &  -  &   2.3 &   1.1 & 1.22E-003 $\pm$ 2.1E-004 & 2.49E-004 $\pm$ 5.6E-005 \\ 
J1757+0531 &     3.4 &    0.06 &    0.08 &   -0.53 & $>$ 11 & $>$ 11 &   2.4 &   3.1 & 1.08E-003 $\pm$ 1.6E-004 & 1.12E-003 $\pm$ 2.4E-004 \\ 
J1800+3848 &     2.2 &    1.00 &    0.82 &    0.36 &  -  &  -  &   0.8 &   1.3 & 9.57E-005 $\pm$ 1.5E-005 & 2.00E-004 $\pm$ 7.5E-005 \\ 
J1812+5603 &     1.8 &    0.47 &    0.46 &    0.04 &  -  & 0.8$\pm$0.5 &   1.0 &   1.7 & 1.75E-004 $\pm$ 4.0E-005 & 5.00E-004 $\pm$ 1.3E-004 \\ 
J1819+3845 &     2.2 &    0.23 &    0.19 &    0.35 & $>$ 11 & 0.2$\pm$0.1 &   2.6 &   2.0 & 1.25E-003 $\pm$ 1.8E-004 & 4.82E-004 $\pm$ 1.2E-004 \\ 
J1832+1357 &     2.3 &    0.33 &    0.31 &    0.12 &  -  &  -  &   1.1 &   1.4 & 1.91E-004 $\pm$ 4.7E-005 & 4.41E-005 $\pm$ 4.7E-006 \\ 
J1839+4100 &     2.4 &    0.07 &    0.10 &   -0.66 &  -  & 0.2$\pm$0.0 &   1.4 &   1.6 & 2.43E-004 $\pm$ 5.3E-005 & 4.43E-004 $\pm$ 1.2E-004 \\ 
J1850+2825 &     5.8 &    1.45 &    1.11 &    0.50 &  -  & 1.5$\pm$10.9 &   1.1 &   3.1 & 9.85E-005 $\pm$ 7.2E-005 & 4.34E-004 $\pm$ 6.8E-005 \\ 
J1905+1943 &     3.2 &    0.21 &    0.26 &   -0.39 & 0.4$\pm$0.2 & 0.5$\pm$1.3 &   1.6 &   2.0 & 4.46E-004 $\pm$ 8.3E-005 & 6.89E-004 $\pm$ 3.4E-004 \\ 
J1919+3159 &     6.5 &    0.11 &    0.11 &    0.03 & $>$ 11 & $>$ 11 &   6.2 &   5.1 & 1.05E-002 $\pm$ 1.2E-003 & 3.23E-003 $\pm$ 1.5E-003 \\ 
J1931+4743 &     5.2 &    0.10 &    0.11 &   -0.12 & 0.0$\pm$0.1 & 0.1$\pm$0.3 &   7.4 &   7.2 & 9.83E-003 $\pm$ 1.7E-003 & 9.04E-003 $\pm$ 2.6E-003 \\ 
J2002+4725 &    14.7 &    0.87 &    0.98 &   -0.22 & 3.4$\pm$2.4 & 1.4$\pm$0.7 &   1.7 &   2.0 & 6.72E-004 $\pm$ 6.9E-005 & 8.22E-004 $\pm$ 1.9E-004 \\ 
J2006+6424 &     4.3 &    0.83 &    0.48 &    1.02 & $>$ 11 & $>$ 11 &   2.2 &   3.3 & 1.03E-003 $\pm$ 1.4E-004 & 1.70E-003 $\pm$ 2.2E-004 \\ 
J2011+7205 &     4.8 &    0.10 &    0.11 &   -0.28 & 1.0$\pm$0.6 & 2.3$\pm$1.6 &   7.1 &   6.4 & 1.02E-002 $\pm$ 8.0E-004 & 1.02E-002 $\pm$ 1.1E-003 \\ 
J2012+6319 &     3.9 &    0.11 &    0.13 &   -0.24 & 0.3$\pm$0.2 & 1.3$\pm$0.5 &   3.9 &   4.9 & 3.16E-003 $\pm$ 5.5E-004 & 5.27E-003 $\pm$ 1.7E-003 \\ 
J2016+1632 &     3.6 &    0.56 &    0.47 &    0.33 & $>$ 11 & $>$ 11 &   2.2 &   4.8 & 8.71E-004 $\pm$ 1.5E-004 & 5.09E-003 $\pm$ 1.8E-003 \\ 
J2113+1121 &     1.9 &    0.06 &    0.07 &   -0.09 & 0.4$\pm$0.0 & 4.6$\pm$2.9 &   5.6 &   7.9 & 5.89E-003 $\pm$ 1.4E-003 & 1.76E-002 $\pm$ 4.2E-003 \\ 
J2116+0536 &     1.4 &    0.18 &    0.21 &   -0.22 & 0.5$\pm$0.0 & 6.0$\pm$1.0 &   2.4 &   4.4 & 1.02E-003 $\pm$ 2.4E-004 & 3.75E-003 $\pm$ 9.6E-004 \\ 
J2123+0535 &     1.4 &    1.86 &    2.06 &   -0.19 &  -  & 10.0$\pm$6.7 &   0.9 &   1.6 & 1.02E-004 $\pm$ 4.4E-005 & 4.49E-004 $\pm$ 1.8E-004 \\ 
J2137+0451 &     1.3 &    0.11 &    0.13 &   -0.30 & 0.6$\pm$0.5 & 1.2$\pm$0.3 &   2.1 &   2.1 & 8.43E-004 $\pm$ 8.6E-005 & 7.02E-004 $\pm$ 2.1E-004 \\ 
J2203+1725 &     1.7 &    1.04 &    0.98 &    0.10 & 0.4$\pm$0.1 & $>$ 11 &   2.0 &   3.7 & 6.20E-004 $\pm$ 2.0E-004 & 2.79E-003 $\pm$ 9.4E-004 \\ 
J2208+1808 &     1.4 &    0.05 &    0.09 &   -1.15 &  -  &  -  &   1.5 &   1.4 & 1.31E-004 $\pm$ 1.1E-004 & 1.66E-004 $\pm$ 7.8E-005 \\ 
J2212+2355 &     2.3 &    1.00 &    0.96 &    0.07 & 0.3$\pm$0.2 & $>$ 11 &   2.5 &   2.3 & 9.72E-004 $\pm$ 1.8E-004 & 1.11E-003 $\pm$ 2.0E-004 \\ 
J2221+1117 &     1.0 &    0.08 &    0.08 &    0.12 & 0.2$\pm$0.2 & 0.2$\pm$0.4 &   2.2 &   2.9 & 7.25E-004 $\pm$ 2.6E-004 & 1.05E-003 $\pm$ 5.8E-004 \\ 
J2237+4216 &     5.1 &    0.20 &    0.23 &   -0.19 & 0.7$\pm$0.6 & 1.2$\pm$1.1 &   3.8 &   5.9 & 3.40E-003 $\pm$ 4.6E-004 & 8.67E-003 $\pm$ 1.7E-003 \\ 
J2241+0953 &     1.4 &    0.53 &    0.60 &   -0.21 &  -  & 0.5$\pm$0.7 &   1.0 &   1.6 & 1.63E-004 $\pm$ 5.9E-005 & 5.14E-004 $\pm$ 2.7E-004 \\ 
J2242+2955 &     2.3 &    0.10 &    0.11 &   -0.27 & 0.1$\pm$0.1 & 6.2$\pm$3.2 &   2.0 &   3.9 & 6.77E-004 $\pm$ 9.1E-005 & 3.09E-003 $\pm$ 7.1E-004 \\ 
J2253+3236 &     3.0 &    0.19 &    0.19 &   -0.03 & 0.1$\pm$0.1 & 0.0$\pm$0.2 &   2.7 &   2.4 & 1.35E-003 $\pm$ 2.3E-004 & 5.90E-004 $\pm$ 2.6E-004 \\ 
J2258+0516 &     0.9 &    0.19 &    0.21 &   -0.19 &  -  &  -  &   1.2 &   1.1 & 2.50E-004 $\pm$ 7.7E-005 & 3.82E-005 $\pm$ 3.0E-005 \\ 
J2304+2710 &     1.3 &    0.12 &    0.10 &    0.27 & 0.2$\pm$0.1 & 0.2$\pm$0.0 &   2.5 &   3.4 & 1.17E-003 $\pm$ 2.1E-004 & 2.17E-003 $\pm$ 5.5E-004 \\ 
J2311+4543 &     3.2 &    0.35 &    0.28 &    0.40 & 0.6$\pm$0.5 & 1.2$\pm$0.8 &   3.3 &   3.6 & 2.33E-003 $\pm$ 3.9E-004 & 3.13E-003 $\pm$ 7.7E-004 \\ 
J2315+8631 &     2.3 &    0.24 &    0.24 &   -0.03 & 0.5$\pm$0.4 & 1.8$\pm$1.3 &   1.5 &   2.8 & 4.10E-004 $\pm$ 4.6E-005 & 1.72E-003 $\pm$ 2.1E-004 \\ 
J2339+0244 &     0.6 &    0.08 &    0.09 &   -0.33 & 0.3$\pm$0.1 & 0.9$\pm$1.2 &   2.3 &   1.6 & 1.04E-003 $\pm$ 2.1E-004 & 3.49E-004 $\pm$ 2.5E-004 \\

\enddata

\tablecomments{(1) IAU name (J2000.0), (2) WHAM H$\alpha$ Intensities (Rayleighs) \citep{haffneretal03}, (3) Mean flux density at 8.4 GHz (Jy), (4) Mean flux density at 4.9 GHz (Jy), (5) Source spectral index, (6) Estimated characteristic timescale of source variability at 8.4 GHz (days), (7) Estimated characteristic timescale of source variability at 4.9 GHz (days), (8) Raw modulation index at 8.4 GHz with no error subtraction (\%), (9) Raw modulation index at 4.9 GHz with no error subtraction (\%), (10) 4-day SF at 8.4 GHz with $D_{noise}$ subtracted, (11) 4-day SF at 4.9 GHz with $D_{noise}$ subtracted.}

\end{deluxetable}